\documentclass[aps,superscriptaddress,twocolumn,prl,nofootinbib]{revtex4-1}
\usepackage[utf8x]{inputenc}
\usepackage{graphicx}
\usepackage{float}
\usepackage{amsmath}
\usepackage{color}
\usepackage[normalem]{ulem}
\usepackage{enumerate}

\def\av#1{\langle#1\rangle}
\usepackage{color}

\begin{document}

\title{The price of a vote: diseconomy in proportional elections} 
\author{Hygor Piaget M. Melo}
\email{Corresponding author: hygor.piaget@gmail.com}
\affiliation{Instituto Federal de Educa\c c\~ao, Ci\^encia e 
Tecnologia do Cear\'a, Avenida Des. Armando de Sales Louzada, 
Acara\'u, Cear{\'a}, Brazil.}
\address{Departamento de F\'isica, Universidade Federal do Cear{\'a},
60451-970 Fortaleza, Cear{\'a}, Brazil.}
\author{Saulo D. S. Reis}
\author{Andr\'e A. Moreira}
\address{Departamento de F\'isica, Universidade Federal do Cear{\'a},
60451-970 Fortaleza, Cear{\'a}, Brazil.}
\author{Hern\'an A. Makse}
\address{Departamento de F\'isica, Universidade Federal do Cear{\'a},
60451-970 Fortaleza, Cear{\'a}, Brazil.}
\address{Levich Institute and Physics Department, City College of New York, New York, NY 10031, USA}
\author{Jos\'e S. Andrade Jr.}
\address{Departamento de F\'isica, Universidade Federal do Cear{\'a},
60451-970 Fortaleza, Cear{\'a}, Brazil.}

\begin{abstract}
\noindent
The increasing cost of electoral campaigns raises the need for 
effective campaign planning and a precise understanding of the 
return of such investment. Interestingly, despite the strong impact 
of elections on our daily lives, how this investment is translated 
into votes is still unknown. By performing data analysis and 
modeling, we show that top candidates spend more money \emph{per 
vote} than the less successful and poorer candidates, a sublinearity 
that discloses a diseconomy of scale. We demonstrate that such 
electoral diseconomy arises from the competition between candidates 
due to inefficient campaign expenditure. Our approach succeeds in 
two important tests. First, it reveals that the statistical pattern 
in the vote distribution of candidates can be explained in terms of 
the independently conceived, but similarly skewed distribution of 
money campaign. Second, using a heuristic argument, we are able to 
predict a turnout percentage for a given election of approximately 
63\%. This result is in good agreement with the average turnout rate 
obtained from real data. Due to its generality, we expect that our 
approach can be applied to a wide range of problems concerning the 
adoption process in marketing campaigns.
\end{abstract}

\maketitle

\twocolumngrid

\noindent
{\bf \large Introduction}\\\\
\noindent
Elections exhibit a complex process of negotiations 
between politicians and voters. The past few decades bore witness to 
a steep increase in the expenditure of political campaigns. Take the 
example of the presidential elections in the US. The 1996 campaigns 
cost contestants approximately \$123~million (corrected for 
inflation) altogether, an amount that escalated to nearly \$2 
billion in 2012~\cite{NYtimes}. Although campaign investments have 
grown, the impact of money into the electoral outcome remains not 
fully understood~\cite{Stratmann05,Holbrook96,Johnston06}, and 
conclusions about it are quite contradictory. In some studies, it 
has been argued that incumbent spending is ineffective, and the 
challenger spending, on the other hand, produces large 
gains~\cite{Jacobson1978,GERBER04,Johnston08}. 
Other studies claim that neither incumbent nor challenger spending 
makes any appreciable difference~\cite{Erikson2000,Hillygus03}, a 
theory that dates back to the 
1940's~\cite{Lazarsfeld1944,Finkel1993}. Yet another group argues 
that both challenger and incumbent spending are 
effective~\cite{Krasno1988}.

Despite the questioning about the effectiveness of political 
campaigns as a whole, the election campaign of President Barack 
Obama in 2012 spent more than 65\% of its money on media, including 
TV and radio air time, digital and printing advertising, and
others~\cite{West2013}. Therefore, the direct contact with voters 
is not only a major factor in campaign planning, but it is believed 
to have relevant impact in succeeding to persuade undecided 
voters~\cite{Esser2004}.

Here we address the problem of how campaign expenditure influences 
election outcome. We start by an extensive analysis of data sets 
from the proportional elections in Brazilian states for the federal 
and state congresses, uncovering a ubiquitous nonlinearity on the 
relation between votes and campaign budget. As we will show, 
candidates  can be gathered into different groups of spenders. One  
group is characterized by candidates with low budget campaign and a 
seemingly uncorrelated number of votes. As the money invested on 
campaign increases, a clear correlation between vote and money 
emerges. Interestingly, in this correlated regime, the top 
candidates are those who spend more in political campaign, but with 
a highly counterintuitive result: the more the candidates spend, the 
less vote per dollar they get.

In Economics, a similar effect in which larger companies tend to 
produce goods at increased per-unit costs is known as 
\emph{diseconomy of scale}. Precisely, the diseconomy of scale makes 
reference to a financial drawback resulting from the increase of the 
production scale. It implies that, above a maximum efficient company 
size, the average cost per unit production increases. In other 
words, above this maximum, the more companies invest to increase in 
size, the less return of such investments they get per produced 
unit. The origin of this type of behavior can be manifold. For 
instance, it has been explained in terms of a systematic increase in 
communication costs~\cite{McAfee1995}, or as a consequence of the 
Ringelmann psychological effect, namely, the tendency for 
individuals to become less efficient when working in larger 
groups~\cite{Ringelmann1913}. To the best of our knowledge, this 
study is the first to report the presence of diseconomy of scale on 
elections.

In order to elucidate the mechanisms responsible for this diseconomy 
in elections, we develop a general model for the negotiations 
between candidates and voters whose solution is compared with 
results from the analysis of electoral data sets. An important 
assumption in our model is that votes are considered to be 
``buyable'', whether they are somehow purchased through direct 
contacts between candidates and voters or, indirectly, through media 
campaigns. In this way, since the amount of financial resources 
$m_i$ effectively represents the main convincing  strength of 
candidate $i$, it also provides an upper bound for the number of 
votes that can be received, when competition among candidates is 
regarded as absent. The potential ability of a candidate to acquire 
votes in this model can be estimated, as a first approximation, in 
terms of the identification of the influential 
spreaders~\cite{Kitsak2010,Morone2015}.

A crucial goal here is to show that the competition between 
candidates is the root cause of the diseconomy of scale observed in 
Brazilian elections, mainly due to the fact that, in a scenario 
without competition, any model prediction will have a tendency to 
overestimate the number of votes of top campaign spenders. Our 
results show that the introduction of competition among candidates 
in the model combined with a simple heuristic argument lead to a 
prediction for the turnout rate of elections that is compatible with
the average value from real data. We obtain this by the assumption 
that campaign planners would make use of financial resources 
considering an equitable division of funds per vote.\\

\noindent
{\bf\large Results}\\\\
\noindent
{\bf Empirical findings}\\
\noindent
Our data analysis is based on real data 
sets acquired from recent proportional elections in Brazil, publicly 
available~\cite{TSE}. These data sets are related to the elections 
for the national lower house and state congress in 2014. Brazilian 
elections represent  a quite general and suitable case study to our 
purposes due to a number of special factors. First, Brazil is a 
large country, both in population and land area. It has the fifth 
population of the world spread across roughly 8.5 million km$^2$ 
(over 3 million mi$^2$). Second, in contrast with executive 
elections, representative elections in Brazil have a large number of 
candidates. Additionally, it is compulsory to vote in Brazil. 
Altogether, these factors lead to a huge data set from a quite
diverse electorate.

\begin{figure}[!h]
\begin{center}
\includegraphics[width=8cm]{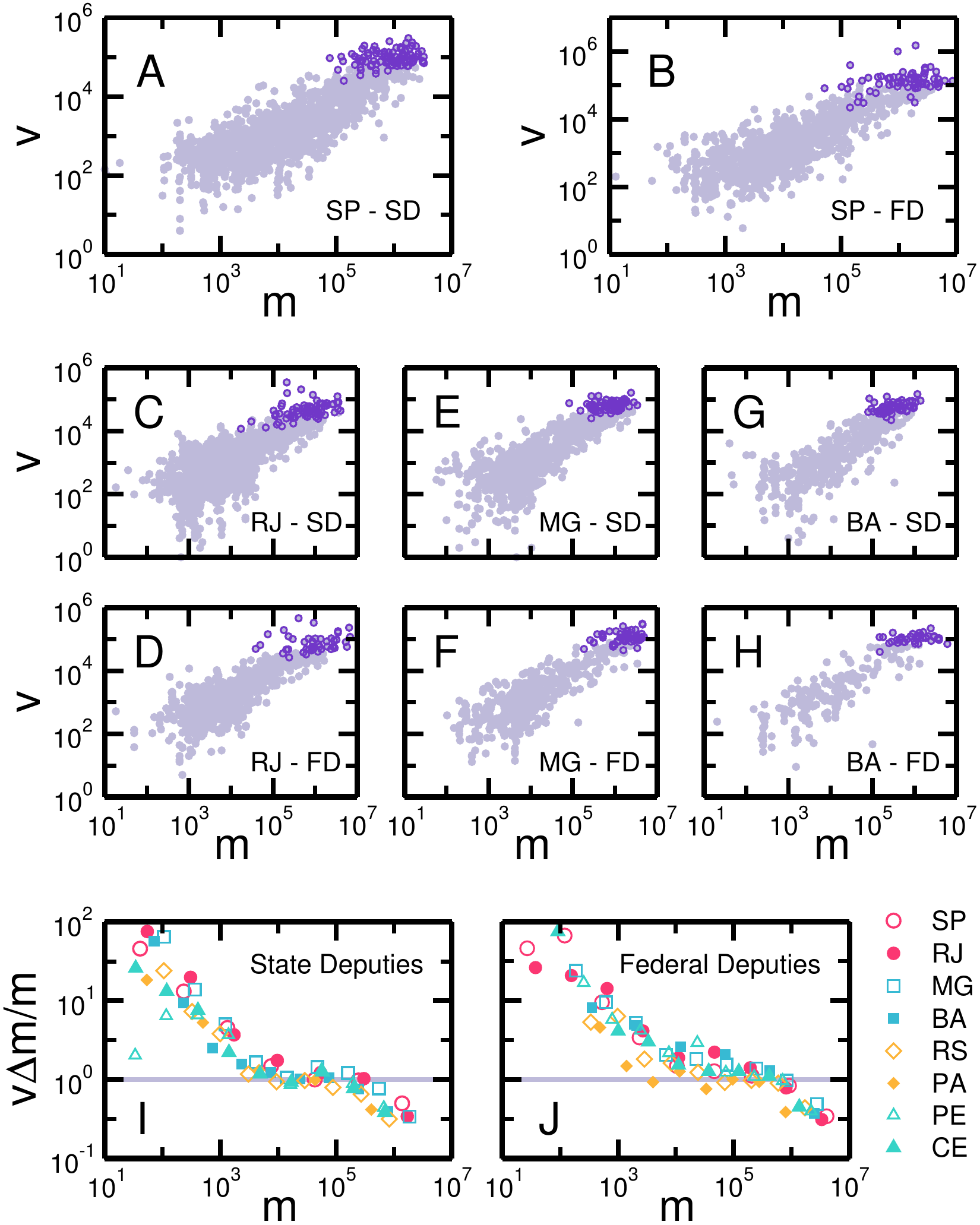}
\caption{\textbf{Scaling relation between number of votes and 
money spent.} The light purple circles show the relation between the 
number of votes and the declared campaign expenditure of each 
candidate in the state (SD) and federal deputies (FD) elections in 
2014 for the four largest states in Brazil: S\~ao Paulo (\emph{A}, 
\emph{B}), Rio de Janeiro (\emph{C}, \emph{D}), Minas Gerais 
(\emph{E}, \emph{F}), and Bahia (\emph{G}, \emph{H}). Despite the 
large fluctuations, there is an unambiguous correlation between 
votes and money. In each panel, the data for elected candidates are 
highlighted in dark purple circles. In order to see the nuances of 
the correlation, we plotted in a normalized relation for (\emph{I}) 
state and (\emph{J}) federal deputies for the eight largest states 
in Brazil. The symbols represent the normalized ratio 
$\langle v\rangle\Delta m /m$, where we first calculate the average 
number of  votes in log-spaced bins along $m$. If we assume a linear 
correlation, the multiplicative constant is $\Delta m=M/n$. The 
normalization provides us a direct observation of the nonlinearity 
in the dependence of votes on money. We see a global sublinear 
behavior, where the wealthier candidates display a lower fraction of 
votes per money.}
\label{fig1}
\end{center}
\end{figure}

\begin{figure*}[ht]
\begin{center}
\includegraphics*[width=17cm]{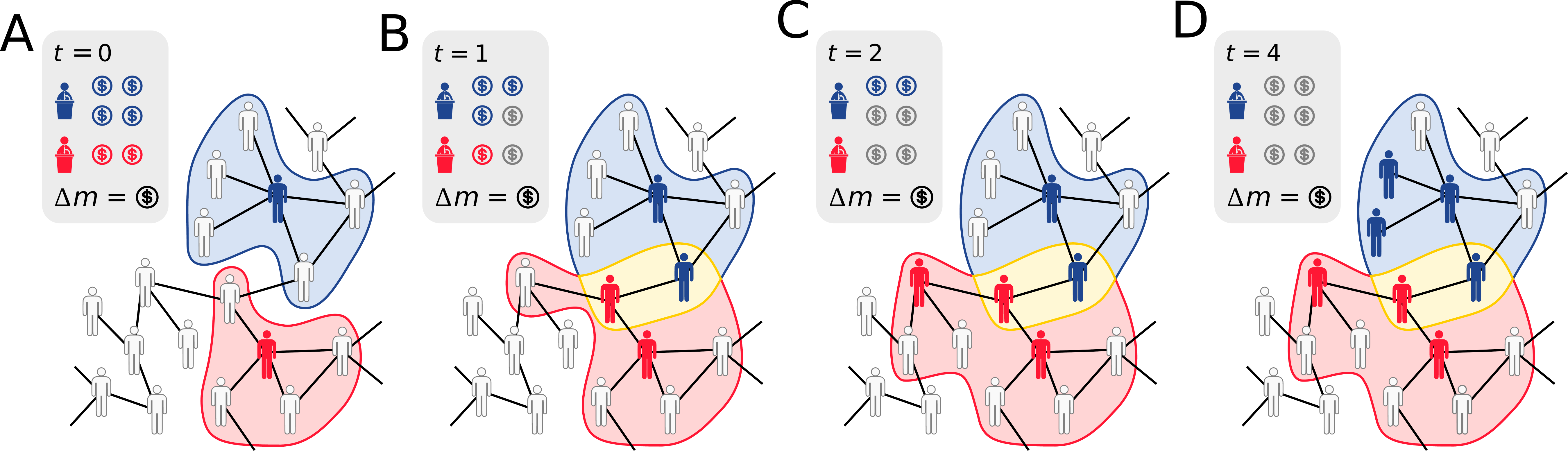}
\caption{\textbf{Model's pictorial sketch.} Here, two candidates 
compete for votes in a social network with undecided (light gray) 
individuals. (\emph{A}) The blue candidate has initially a budget of 
4 (four) monetary units, while the campaign for the red candidate 
has a budget of 2 (two) monetary units. In this example, both 
candidates also has one decided voter at the beginning of the 
process, namely, the blue voter at the top part of the network	and 
the red one at the bottom part of the network. The highlighted 
regions enclose the initial voter each campaign has and the 
acquaintances of each initial voter. These regions represent 
operational areas the campaigns will act at the next time step in 
order to convince undecided voters to vote for their associated 
candidates. (\emph{B}) One undecided voter inside each operational 
area is randomly chosen, becoming then decided voters. Accordingly, 
each budget campaign decreases by the amount of $\Delta m=\$$. As a 
consequence of new decided voters, the operational areas grow, and 
since two acquaintances become voters of different candidates, the 
operational areas of the campaigns now overlap. This region where 
both campaigns can act is represented in yellow. (\emph{C}) While 
the red candidate's campaign chose an undecided voter, increasing 
its operational area, blue campaign ineffectively spends money on a 
decided voter. (\emph{D}) At the end of the campaign process, when 
all campaigns run out of funds, the campaign with larger initial 
budget ends the process with more adopters, but his/her campaign is 
less efficient, resulting in a diseconomy of scale.}
\label{fig2}
\end{center}
\end{figure*}

We start by assembling the data sets on the entire electoral outcome 
and campaign expenditure of candidates from all 26 Brazilian states. 
Figure~1 displays the number of votes $v$ versus the 
declared campaign expenditure $m$ of each candidate for the top 4 
Brazilian states in terms of population, namely, S\~ao Paulo 
(Figs.~1\emph{A} and~1\emph{B}), Rio de Janeiro 
(Figs.~1\emph{C} and~1\emph{D}), Minas Gerais 
(Figs.~1\emph{E} and~1\emph{F}) and Bahia 
(Figs.~1\emph{G} and~1\emph{H}). As depicted in 
Fig.~1, the clouds of points are neatly correlated and 
follow a clear trend. This trend is observed in all representative 
elections for all Brazilian states (see Supporting Information 
Section I). 

To extract the main relationship between $v$ and $m$, we average the 
number of votes in log-spaced bins along $m$, which provides an 
estimation for the empirical relation of  $\av{v}$ as a function of 
$m$. In order to plot results for different states in the same 
figure, we perform a scale transformation on $\av{v}$ by supposing 
simple linear relation $\av{v}=c\times m$, where $c$ is a 
characteristic constant of a given election. If we define the 
average price of a vote as $\Delta m=\sum_i m_i/\sum_iv_i$ and 
suppose that it is roughly uniform across candidates, it is easy to 
see that $c=1/\Delta m$. Here, $v_i$ is the number of votes of 
candidates $i$. If the relation  between votes and money 
is linear, then the plot of $\av{v}\times(\Delta m/m)$ should be a 
constant function of $m$ with value close to $1.0$.

In Figs.~1\emph{I} and 1\emph{J}, we plot 
$\av{v}\Delta m/m$ as a function of $m$ for the state legislative 
assembly and federal congress elections, respectively, for the year 
of 2014 and for the eight most populated states in Brazil. The 
result shows a consistent nontrivial dependence of votes on money 
spent in campaign. For small values of $m$, we observe a rapidly 
decrease of $\av{v}\Delta m/m$. For intermediate expenditures in the 
range R\$10,000~$<m<$~{\rm R}\$100,000\footnote{At election day, 
October 5th, 2014, the exchange rate between the Dollar and the 
Brazilian Real was ${\rm R}\$~2.4266 = \$1$.}, we observe  an 
apparent linear dependence of $v$ with respect to $m$. Finally, for 
$m>$~{\rm R}\$100,000, a noticeable departure from linearity is 
observed, that is, wealthier candidates need a disproportionately 
large amount of money to obtain a single vote as compared with less 
successful candidates within the same range of financial 
resources.\\

\noindent
{\bf A general model for the price of a vote}\\
Here, we propose a general model for the price of a vote. We 
consider an electoral process composed of two separate groups of 
individuals, candidates and voters. All $s$ candidates can compete 
for the vote of all $n$ voters, and each candidate $i$ has a limited 
amount of money $m_i$ to spend on their campaigns. Thus, if at a 
given time $m_i=0$, the candidate becomes unable to compete for 
voters anymore. Here we assume that candidates can only conquer a 
single vote at a given time step and that voters, once they reach a 
decision, cannot change their minds anymore. As compared to the case 
of plurality elections, the last assumption is readily justifiable 
for proportional elections since, in this case, candidates do not 
compete directly for the same seat. As a consequence, voters do not 
feel compelled to rethink their decisions. In this way, because it 
is not possible to know if a voter reached a decision or not, 
campaigns can spend money on already decided voters, leading to 
ineffective use of financial resources.

\begin{figure*}[!t]
\begin{center}
\includegraphics[width=11cm]{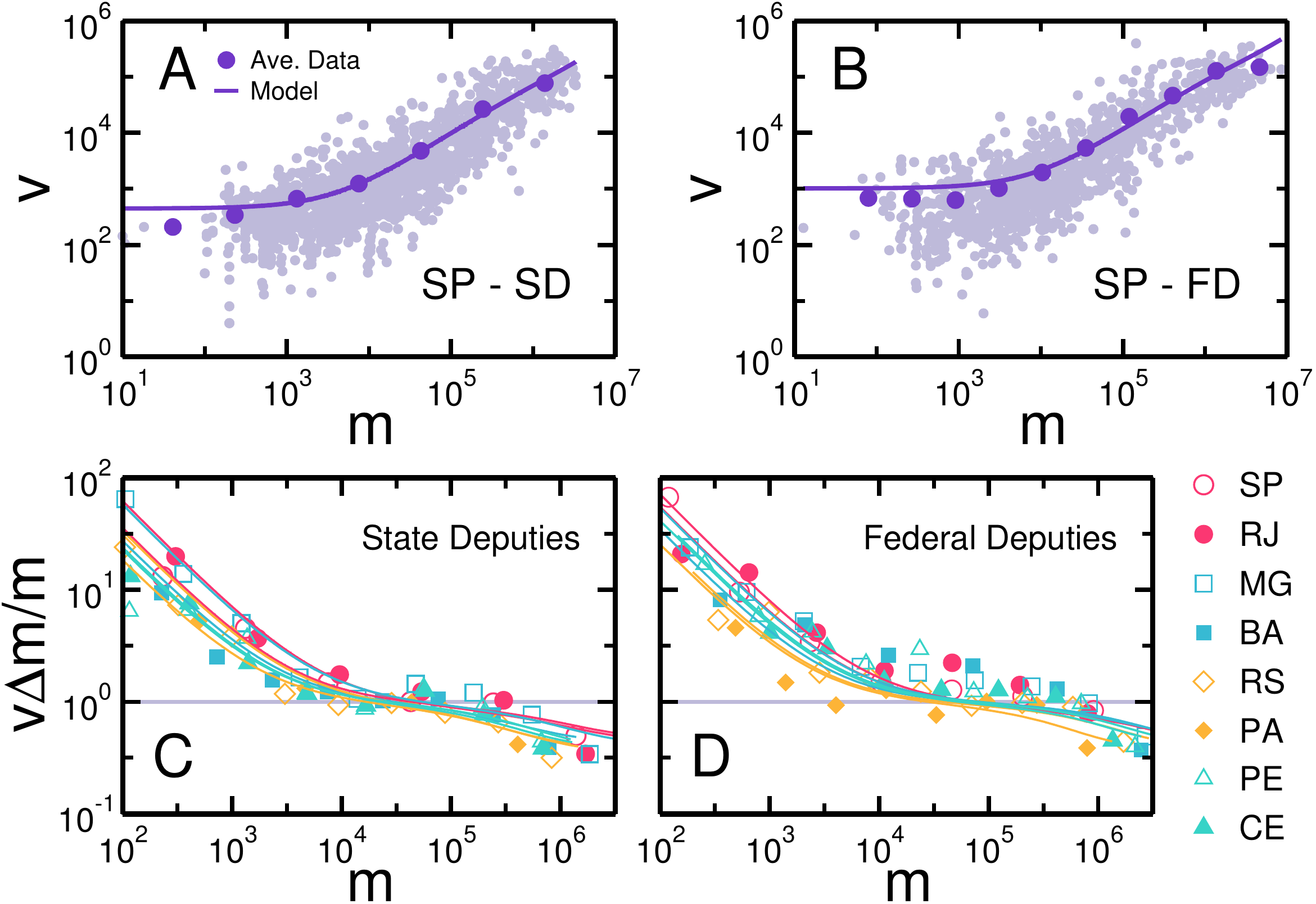}
\caption{\textbf{Modelling the nonlinear scaling.} In order to 
verify if our model correctly fits the data, we show in (\emph{A}) 
and (\emph{B}) the S\~ao Paulo election for state and federal 
deputies in 2014, respectively. Each light purple circle corresponds 
to one candidate and the dark purples circles are the average number 
of votes in log-spaced bins along $m$. We see that our model shows a 
good agreement with the average behavior for all the money spectrum. 
In (\emph{C})  and (\emph{D}) we perform the same normalization 
process as in Fig.~2 but now with $\Delta m$ is estimated 
using Eq.~\ref{eq3}. Each solid line shows the solution of 
our model. Despite its simplicity, our model features all nonlinear 
regimes seen in the data, which corroborates our theory that the 
inefficiency of wealthier candidates are mainly due to 
competition.}
\label{fig3}
\end{center}
\end{figure*}

A pictorial description of the model is presented in 
Fig.~2. On a social network with undecided voters, 
represented by light gray individuals, two candidates start their 
campaigns with an initial amount of money $m$ and one single decided 
voter. This initial seed is represented in Fig.~2\emph{A} 
by the blue and red individuals. The regions highlighted in blue and 
red represent the operational areas of the campaigns, enclosing the 
group of voters to whom the campaigns will spend money in order to 
turn undecided voters into decided voters. As depicted in 
Fig.~2\emph{B}, at each time step each campaign chooses one 
voter inside its operational areas. If the chosen individual is an 
undecided voter, she/he becomes a decided voter. 
Accordingly, the overall campaign money is decreased by an amount of 
$\Delta m$. If the chosen voter is already a decided voter, as 
depicted in Fig.~2\emph{C}, the campaign budget is also 
decreased by $\Delta m$, but the voter's decision remains unchanged. 
We repeat this procedure until all campaigns run out of funds. In 
Fig~2\emph{D}, we show a typical example of a competition 
for votes between two candidates during the electoral process 
described by our model. Although the candidate with the larger 
initial budget receives more votes at the end of the election, due 
to ineffective spending, the campaign of the poorer candidate is, in 
fact, more efficient.

In order to represent the reach of the traditional and social 
medias, as a first approximation, we apply this model on a complete 
graph, so that the time evolution of the number of votes of a 
given candidate $i$ can be written as 
\begin{equation}
\frac{dv_i}{dt}=\left(1-\frac{S(t)}{n}\right) [m_i(t)>0],
\label{eq1}
\end{equation}
where $S(t)=\sum_{i=0}^{s} v_i$ is the total number of decided 
voters at time $t$, and $\left[m_i(t)>0\right]$ is the Iverson 
bracket, which is $1$ if the condition inside the brackets is 
satisfied,  and $0$  otherwise. The right-hand side of the 
Eq.~\ref{eq1} is the probability of candidate $i$ to choose 
an undecided voter at time $t$. Equation~\ref{eq1} explicitly 
requires a definition for the rate of money expenditure, $dm_i/dt$, 
which determines the gradual decrease in financial resources of 
candidate $i$. As simplifying assumptions, we consider that the 
amount of money spent during the campaign decreases linearly,
$dm_{i}/dt=-\Delta m_{i}$, and that this constant rate is the same 
for all candidates, $\Delta m_{i}=\Delta m$, $\forall i$. 

The probabilistic feature of Eq.~\ref{eq1} is central to 
confirm our hypothesis that electoral outcome is an output of 
campaign expenditure due to a competition process. This is shown 
here by first considering the case without competition, where 
$s\ll n$. Also, we assume that $n\Delta m\gg m_i$ for all $i$, so 
that the candidate with the highest amount of funds do not have 
enough money to reach out the whole network. By doing so, it is 
unlikely that the extent of the candidates' campaigns overlap, and 
therefore, a candidate would not waste her/his campaign money on a 
decided voter of another candidate.  As a consequence, since  the 
probability of candidate $i$ to conquer an undecided voter is not 
affected by another campaign, $S(t)$ can be replaced 
by $v_{i}$ in Eq.~\ref{eq1}, leading to an uncoupled system 
of differential equations, whose solution is given by,
\begin{equation}
v_i=n-(n-v_{0,i})e^{-m_i/n\Delta m},
\label{eq2}
\end{equation}
where $v_{0,i}$ is the initial number of votes of candidate $i$. 
Since $n\Delta m\gg m_i$, and assuming that $(n-v_{0,i})\approx n$, 
by expanding the exponential and taking its first order 
approximation, we can write the number of votes as
$v_i\approx v_{0,i}+m_i/\Delta m$ . As we discuss next, this simple 
model do not suffice to explain the whole complexity of the relation 
between $v$  and $m$. The first two regimes presented in
Fig.~2 can be understood in therms of this approximation. 
For the regime of low $m$, where the experimental data do not  
exhibit a clear correlation, the candidates start the race with 
$v_{0,i}$ votes. Since they cannot afford a long run and/or a large 
expenditure, their final performance fluctuates around the initial 
value $v_{0,i}$, which depends on different factors, such as free 
volunteer engagement. As campaign money increases, the linear part 
overcomes the initial number $v_{0,i}$, and a linear regime emerges. 
However, in the scenario without competition, the linear behavior 
remains at large $m$.

We now consider the competition between candidates as a possible 
cause for the transition from linear to sublinear regime. 
Disregarding all previous simplifying assumptions and integrating 
Eq.~\ref{eq1}, we find
\begin{equation}
v_i=v_{0,i}+\frac{m_i}{\Delta m} - \frac{1}{n\Delta m}\int_0^{m_i}S(m')dm',
\label{eq3}
\end{equation}
where the integration of the Iverson bracket over time gives the 
total time candidate $i$ has to perform her/his campaign, 
$m_i/\Delta m$, and we used $dm'/dt'=-\Delta m$ to change the 
variable of integration on the last term.

It is possible to find a differential equation for $S(m')$ by taking
Eq.~\ref{eq1} and summing over $i$. After solving it for 
$S(m')$ and integrating the last term of Eq.~\ref{eq3} (see 
Supporting Information Section II for details of the analytical 
solution), we find a set of nonlinear coupled equations that must be 
solved, candidate by candidate, following an increasing order of 
$m_i$ values. As a consequence, the number of votes of candidate $i$ 
depends on the whole distribution $P(m)$ through the integral term 
in Eq.~\ref{eq3}.

Equation~\ref{eq3} has a simple interpretation. As in the case 
without competition, all candidates begin their run with an initial 
number of votes, and those with sufficient money to keep running 
enter in a linear regime controlled by the rate $\Delta m/m$. 
Nonetheless, as we will see next, candidates with sufficient 
campaign funds may start to waste their money on decided voters, a 
behavior that is substantiated by the presence of $S(m')$ in the 
last term of Eq.~\ref{eq3}, which encloses the competition 
dynamics. We consider this collective  influence of the total 
financial resources from all candidates during the campaign as an 
important result, since it  provides a bridge between campaign
expenditure and electoral outcome, which is the basis of the 
remaining results that follows.

In order to obtain a solution for the model, we use as inputs the 
money $m_i$ of each candidate $i$, obtained from data, the total 
number of voters $n$, an initial number of votes $v_0$, and an 
estimated value for $\Delta m$. For all candidates, we define 
$v_{0}$  as the average number of votes of candidates with less 
then R\$1,000. The parameter $\Delta m$ is calculated as a function 
of the turnout rate $T=S_f/n$, where $S_{f}=\lim_{t\to\infty}S(t)$ 
is the total number of votes at steady state. We can therefore write  
the final fraction of votes as
\begin{equation}
T=1-e^{-M/(n\Delta m)},
\label{eq4}
\end{equation}
where $M=\sum_i m_i$ is the total amount of money in the campaign 
process. Therefore, we estimate $\Delta m$ using Eq.~\ref{eq4} such 
that the total number of votes fits the turnout election data.

\begin{figure*}[!t]
\begin{center}
\includegraphics*[width=14cm]{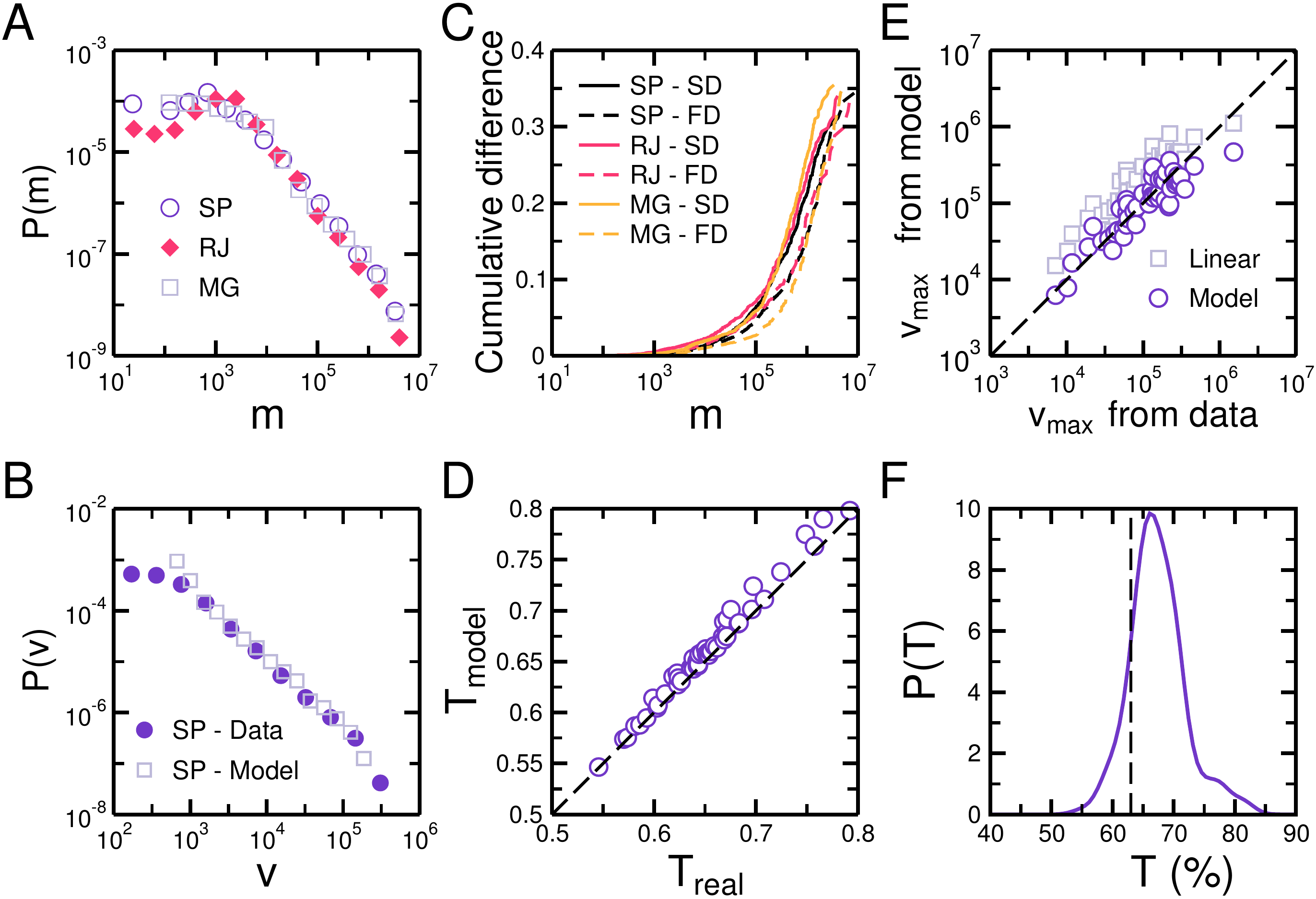}
\caption{\textbf{Analytical results of the model.} In order to 
derive the distribution of votes, our model takes as input the 
distribution of money. (\emph{A}) We see that the distributions of 
money for the state deputies in S\~ao Paulo (SP), Rio de Janeiro 
(RJ) and Minas Gerais (MG) reveal long tails characteristic. 
(\emph{B}) We can now compare the actual distribution (circles) of 
votes, $P(v)$, with the ones obtained by our model (squares) for the 
election of state representatives in S\~ao Paulo. Clearly, we can 
see that our model have a good agreement with the data showing that 
the universal long tail characteristic of $P(v)$ is a direct 
consequence of the money as an input for dynamical competition 
process. (\emph{C}) Relative difference between the cumulative 
distributions of the predictions for the model with and without 
competition. Here we show the results for the elections of state 
deputies (SD) and federal deputies (FD) for S\~ao Paulo (SP), Rio de 
Janeiro (RJ) and Minas	Gerais (MG). No noticeable difference between 
both approaches can be observed for the region of 
$m<{\rm R}\$10,000$. However, for the region of top spenders 
candidates ($m>{\rm R}\$1,000,00$), those who get elected, the 
relative difference is drastic, varying from $30\%$ to $40\%$. 
(\emph{D}) By solving our model, as expressed in Eq.~\ref{eq1}, we 
calculate the expected number of votes that each candidate should 
have for an election. The total number of votes of that election 
divided by the number of voters $n$ is defined as the turnout ratio 
$T$. For all 56 parliamentary elections in 2014, we compared our 
estimation of the turnout ratio, $T_{\rm model}$, and the data ratio 
$T_{\rm data}$. The dashed line represents what would be the perfect 
agreement, $T_{\rm real}=T_{\rm model}$. As can be seen, the 
simulations (circles) exhibit a good agreement with the data. 
(\emph{E}) We can also pick up the candidate with the largest number 
of votes $v_{\max}$ and see how our model estimates this value. As 
depicted, we see that the competition model (circles) better 
estimates $v_{\rm max}$ when compared with the linear model 
(squares), which always overestimates it. The non-parametric 
histogram of $T$ shown in (\emph{F}) for the election of 2006, 2010, 
and 2014 reveals an average turnout value of approximately 67\, 
which is consistent with our heuristic estimation of 
$T=1-e^{-1}\approx63\%$ (vertical dashed line).}
\label{fig4}
\end{center}
\end{figure*}

The results of the election in S\~ao Paulo state for state and 
federal deputies in 2014 are shown in Figs.~3\emph{A} 
and~3\emph{B}, respectively. As depicted, the predictions 
of our model (solid line) are in good agreement with the average 
values of the number of votes for different classes of candidates in 
terms of fund raising. Note that no fitting parameters are necessary 
for this comparison. For $m<$~{\rm R}\$1,000, our model exhibits a 
constant behavior, capturing the uncorrelated nature of the data. 
Additionally, for $m>$~{\rm R}\$1,000, an evident correlation 
between votes and money is present. This is better visualized when 
we plot in Figs.~3\emph{C} and~3\emph{D} the 
normalized ratio $\langle v\rangle\Delta m/m $ for the eight most 
populated Brazilian states. Here, the symbols represent the data 
average and the lines show the solution of our model for each state 
identified by color. For small and large values of $m$, we see that 
our model exhibits a clear deviation from a linear behavior. In 
other words, besides exhibiting this deviation for 
$m<$~{\rm R\$}1,000, a clear sublinearity is present for 
$m>$~{\rm R\$}100,000. Under the perspective of our model, the 
observed diseconomy of scale is a direct consequence of the 
competition among candidates (see Supporting Information Section 
III for a statistical comparison between our model with competition 
and the linear model without competition). 

Social networks are known to display the small-world phenomenon, 
where the typical network distance between two individuals, $\ell$, 
is rather small when compared to the system size, 
$\ell\sim \log N$~\cite{Milgram67,Barabasi2016}. Our analytical 
solution on a complete graph works as a first approximation of such 
complex social network structure. In order to compare our model with 
a more realistic one, we apply the dynamics presented on 
Fig.~2 on a random graph~\cite{Barabasi2016} (see 
Supporting Information Section IV). We found a good agreement 
between the solution on a complete graph model and the numerical 
simulation results obtained with a random graph model.\\

\noindent
{\bf Frequency distribution of votes}\\
One of the first empirical 
investigations concerning Brazilian elections was carried out to 
determine the distribution $P(v)$ of the number of candidates 
receiving $v$ votes~\cite{Costa99,Costa2003,Moreira06,Loreto09}. 
Since then, several other studies have been devoted to elucidate the 
origin of the anomalous behaviors of $P(v)$ for other countries as 
well as to propose mathematical models that can provide some insight 
on the social and political mechanisms responsible for this 
statistical behavior~\cite{Moreira06,Calvao2015,Fortunato2007}. In 
our modeling approach, however, the distribution of votes emerges 
as a natural outcome of the distribution of financial resources 
$P(m)$. As shown in Fig.~4\emph{A}, the distribution 
$P(m)$ calculated for state deputies of three different states in 
Brazil can all be described in terms of a power-law type of decay 
extending over a region of approximately six orders of magnitude. 
Using those distributions as inputs, we determine $P(v)$ for each 
one of those elections. In Fig.~4\emph{B} we compare the 
empirical votes distribution for the state of S\~ao Paulo with the 
one obtained by our model, which reproduces correctly the empirical 
distribution of votes among candidates, $P(v)$, for over two orders 
of magnitude (see Supporting Information Sec. V for results 
concerning the states of Rio de Janeiro and Minas Gerais). This 
implies that the observed non-Gaussian long tail form has its origin 
in the heterogeneous aspect of the distribution of campaign 
resources, regardless of the intricate social network and 
information dynamics behind the electoral process.\\

\noindent
{\bf Model validation}\\
To highlight the effect of the sublinearity on forecasting an 
election, we compute the relative difference between the cumulative 
vote distribution predicted by the linear model without competition 
and the one predicted by the model with competition. As shown in 
Fig.~4\emph{C}, for state congress election in the top 
three populated Brazilian states, namely, S\~ao Paulo, Rio de 
Janeiro, and Minas Gerais, no significant difference is noticed 
between the two predictions for campaigns of low expenditure. 
However, for electoral campaigns that invested more than 
${\rm R\$}10,000$, a substantial discrepancy between predictions can 
be noticed. For this region of top spenders, the cumulative 
difference can be drastic, going above $30\%$ in some cases.

We confirm the validity of our model by comparison with data from 
the 2014 state and federal deputy elections that took place 
simultaneously in the 26 states of Brazil. As shown in
Fig.~4\emph{D}, where each point corresponds to an election 
in a given state, the model results for the turnout rate $T$, as 
provided by Eq.~\ref{eq4}, are compatible with the observed 
data. This agreement only confirms the self-consistency of our 
approach, since Eq.~\ref{eq3} has been used to estimate the 
parameter $\Delta m$. The predictive capability of the model can be 
effectively tested by comparing its estimate with real data for the 
largest number of votes obtained by a candidate in each election, 
$v_{max}$.  As shown in Fig.~4\emph{E}, while the results 
of our model (circles) gather around the identity line, 
demonstrating good quantitative agreement with real data, the linear 
approximation model, $v_{max}\approx v_{0} + m_{max}/\Delta m$ 
(squares), clearly overestimates the values of $v_{max}$.

At this point, we show that our theoretical framework allows for a 
forecast of the turnout ratio $T$, if the following assumptions are 
considered:  (i) the candidates have knowledge of the total amount 
of resources $M$ during the campaign, and ({\it ii}) $\Delta m=M/n$, 
which corresponds to the most simple and equitable division of 
votes. As matter of fact, this last point is equivalent to assume 
that a complete turnout can be achieved, namely, $T=100\%$, as in 
the case without competition. In other words, the candidates devise 
their strategy presupposing that they will obtain the maximum 
possible number of votes, therefore disregarding the competition 
among them. This heuristic argument leads to a fraction of valid 
votes, $T=1-e^{-1} \approx 0.63$.  As shown in 
Fig.~4\emph{F}, the histogram of the number of total valid 
votes for all Congress elections in the years 2006, 2010 and 2014 
indicates an average turnout value of 0.67, which is in close 
agreement with our model prediction. Finally, we also tested our 
theoretical approach by applying the principle of maximum 
entropy~\cite{Jaynes1957} and found that the statistical dispersion 
of the model is consistent with real data from elections (see 
Supporting Information Sec. VI).\\

\noindent
{\bf\large Discussion}\\\\
As a result of the competition between candidates in real elections, 
the nonlinear relation between $v$ and $m$ obtained here can 
complement other statistical analyses for political campaign and 
electoral 
outcome~\cite{Klimek2012,Borghesi2010,Nuno2010,Borghesi2012}. These 
analyses enable the detection of a number of statistical patterns of 
electoral processes, such as the relations between party size and 
temporal correlations~\cite{Andresen2008}, the relations between the 
number of candidates and voters~\cite{Mantovani2011}, and the 
distribution of 
votes~\cite{Costa99,Calvao2015,Fortunato2007,Fortunato2013}. 
Our approach goes beyond the examination of statistical patterns by 
providing a theoretical framework that clarifies a number of key 
issues on the economical features of electoral campaigns. First, we 
proposed a simple modeling framework, whose analytical  solution is 
statistically consistent with extensive data relating financial 
resources of political  and electoral outcomes. Interestingly, the 
same model also provides estimates for the distribution of votes 
among candidates and the electorate turnout rate that are in good 
agreement with real data.

A close inspection of the campaign data investigated here reveals a 
ubiquitous nontrivial relation between $v$ and $m$ for all elections 
investigated. More precisely, we observed that this relation is an 
unambiguous sublinear correlation between the money spent by 
candidate and her/his number of votes $v$, specially for the top 
spender candidates, indicating that the electoral process works in a 
state of diseconomy of scale. To explain this behavior in the 
campaign economy, we propose a general model for marketing where 
candidates compete with each other and must spend their money in 
order to get votes. Despite its simplicity, the model proves capable 
of reproducing the  complexity of the dependence of $v$ with respect 
of $m$. This good agreement makes  our model a possible alternative 
to study other aspects of human collective behavior involving, for 
example, diffusion of innovation and decision-making, such as the 
competition in market share where companies invest in advertising 
for products.\\

\noindent
{\bf Acknowledgements}\\
We thank the Brazilian agencies CNPq, CAPES, FUNCAP, and the 
National Institute of Science and Technology for Complex Systems 
(INCT-SC) in Brazil for financial support, and the support by Army 
Research Laboratory Cooperative Agreement Number W911NF-09-2-0053 
(the ARL Network Science CTA) in the US.\\

\noindent
{\bf Author contributions}\\
\noindent
All authors contributed to all parts of the study.\\

\noindent
{\bf Additional information}\\
\noindent
Supplementary information accompanies this manuscript.\\

\noindent
{\bf Competing interests}\\
\noindent
The authors declare no competing interests.

\pagebreak
\onecolumngrid
\begin{center}
\textbf{\large Supplementary Information: The price of a vote: diseconomy in proportional elections}
\end{center}
\setcounter{equation}{0}
\setcounter{figure}{0}
\setcounter{table}{0}
\setcounter{page}{1}
\makeatletter
\renewcommand{\theequation}{S\arabic{equation}}
\renewcommand{\thefigure}{S\arabic{figure}}
\renewcommand{\bibnumfmt}[1]{[S#1]}
\renewcommand{\citenumfont}[1]{S#1}

\section{The Data}

\subsection{Data Description}

In the main text we investigate the effect of the investment of 
candidates on campaign thanks to the available data containing the 
total donation received by and the expenses of each candidate. 
We analyze Brazilian elections for two different kinds of 
legislators, more specifically, the federal and state deputies. 
Their function is to legislate in the unicameral system of each 
Brazilian state. The federal deputies are representatives in the 
chamber of deputies of the national Congress. They are also elected 
for a four year term by a proportional system. The number of elected 
federal deputies is proportional to the population of each one of 
the 26 states. The data is available at the website of the Brazilian 
Federal Electoral Court~\cite{S_TSE}. By force of law, each candidate 
must provide a detailed description of his/her campaign expenditure 
with specific informations such as the value, date and type of 
expense. All this information can be accessed by the public, however 
in order to know the total cost of the campaign and the number of 
votes of each candidate, it is necessary to process the database 
computationally. In Tables~\ref{States_table_Dep_fed} and 
\ref{States_table_Dep_est}, we show a detailed description of the 
data for each state. State deputies are local representatives 
elected for a four year term by a proportional system. 

\subsection{Results for all States}
\begin{figure*}[!h]
\begin{center}
\includegraphics*[width=0.8\columnwidth]{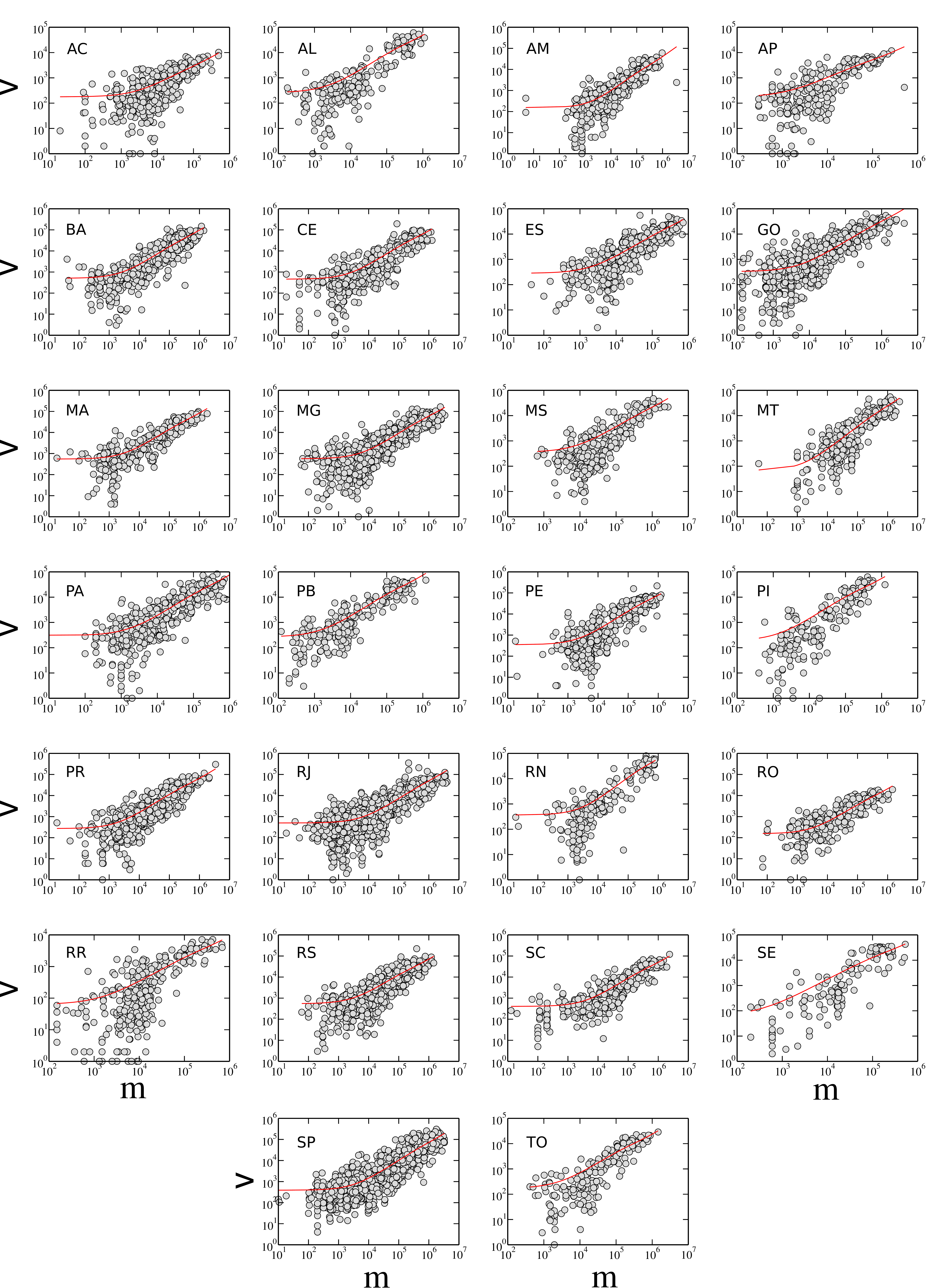}
\caption{\textbf{Modeling the nonlinear scaling for state deputies 
in all federal states.} We show how the model fits the data of state 
deputies election for all states in alphabetic order (AC: Acre, AL: 
Alagoas, AM: Amazonas, AP: Amap\'a, BA: Bahia,  CE: Cear\'a, ES:
Esp\'irito Santo, GO: Goi\'as, MA: Maranh\~ao, MG: Minas Gerais, MS: 
Mato Grosso do Sul, MT: Mato Grosso, PA: Par\'a, PB: Para\'iba, PE: 
Pernambuco, PI: Piau\'i, PR: Paran\'a, RJ: Rio de Janeiro, RN: Rio 
Grande do Norte, RO: Rond\^onia, RR: Roraima:, RS: Rio Grande do 
Sul, SC: Santa Catarina, SE: Sergipe, SP: S\~ao Paulo, TO: 
Tocantis). Each gray circle represents the data for one candidate 
and the red line is the result of the analytical model. We see that 
the model shows a good agreement with the average behavior for all 
states. }
 \label{figS1}
\end{center}
\end{figure*}

\begin{figure*}[!h]
\begin{center}
\includegraphics*[width=0.8\columnwidth]{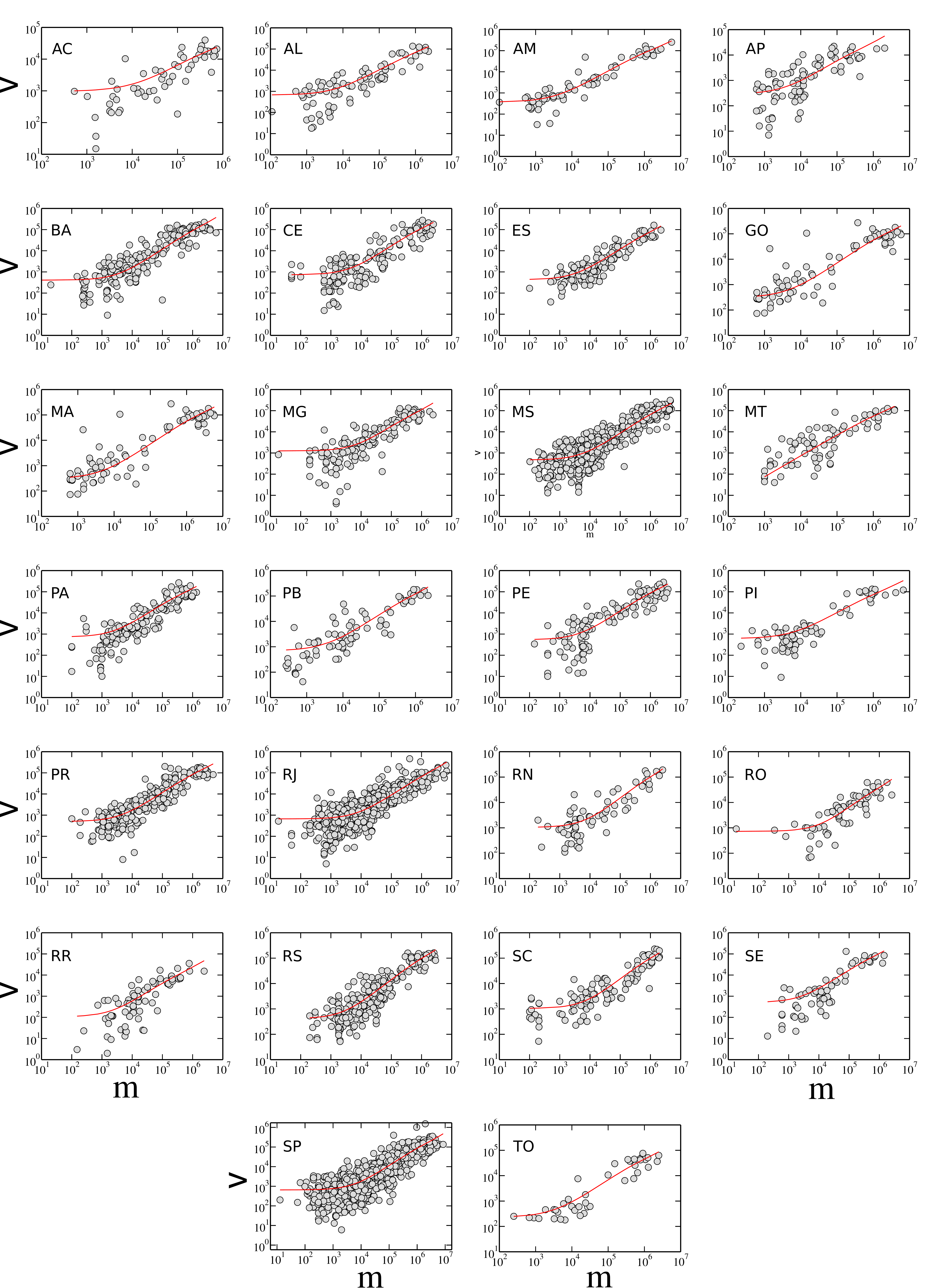}
\caption{\textbf{Modeling the nonlinear scaling for federal deputies 
in all federal states.} We show how the model fits the data of 
federal deputies election for all states in alphabetic order (AC: 
Acre, AL: Alagoas, AM: Amazonas, AP: Amap\'a, BA: Bahia,  CE: 
Cear\'a, ES: Esp\'irito Santo, GO: Goi\'as, MA: Maranh\~ao, MG: 
Minas Gerais, MS: Mato Grosso do Sul, MT: Mato Grosso, PA: Par\'a, 
PB: Para\'iba, PE: Pernambuco, PI: Piau\'i, PR: Paran\'a, RJ: Rio de
Janeiro, RN: Rio Grande do Norte, RO: Rond\^onia, RR: Roraima:, RS: 
Rio Grande do Sul, SC: Santa Catarina, SE: Sergipe, SP: S\~ao Paulo, 
TO: Tocantis). Each gray circle represents the data for one 
candidate and the red line is the result of the analytical model. We 
see that the model shows a good agreement with the average behavior 
for all states. } 
\label{figS2}
\end{center}
\end{figure*}

Here we summarize the results of our model for the election in 2014 
of state and federal deputies in each Brazilian state. 
Figure~\ref{figS1} shows the data obtained for state deputies 
election and Fig.~\ref{figS2} shows data for the federal deputies
election. 

\section{Analytical Solution}
\subsection{Calculation of the expected turnout rate $T$}
Following from Eq.~(1) in the main text and summing over $i$, we 
can find a differential equation for the decided number of voters
$S$, which reads
\begin{equation}
\frac{dS}{d t}=\left(1-\frac{S(t)}{n}\right)r(t),
\label{eq11}
\end{equation}
where $n$ is the total number of voters, and
$r(t)=\sum_i [m_i(t)>0] $ is the number of candidates who 
still have money at instant $t$, which depends solely on the 
distribution of money. After integrating Eq.~(\ref{eq11}), we find 
that
\begin{equation}
S(t)=n-(n-S(0))\exp\left(-\frac{1}{n} \int_0^t r(t')dt' \right).
\label{eq12}
\end{equation}
This equation enables us to compute the expected turnout 
rate $T$ of the election as a function of the average price of a 
vote $\Delta m$, the total money $M$, and $n$. 
To compute $T$, it is necessary to take the limit $t\to \infty$, 
first. At this limit, we are able to compute the value where $S$ 
saturates. Then, we can define $T$ as
\begin{equation}
T=\frac{1}{n}\lim_{t\to\infty}S(t).
\label{eq13}
\end{equation}
In order to compute the integral in Eq.~(\ref{eq12}) at this limit, 
we recall from the main text that $dt=-dm/\Delta m$.  Then, the 
integral becomes
\begin{equation}
\lim_{t\to\infty}\int_0^t r(t')dt'=\int_0^\infty \sum_{i=0}^{N_c} [m_i(m')>0]  dm'/\Delta m,
\label{eq14} 
\end{equation}
where $N_c$ is the total number of candidates. After commutating the 
summation with the integral, and integrating the Iverson's 
bracket over $m'$, we find that
\begin{equation}
\lim_{t\to\infty}\int_0^t r(t')dt'= \frac{M}{\Delta m},
\label{eq15}
\end{equation}
which leads to
\begin{equation}
T = 1-e^{-M/(n\Delta m)}.
\label{eq16}
\end{equation}

Figure~\ref{figPol_5}A shows the turnout rate $T$ as a function of 
$\Delta m$ computed from Eq.~(\ref{eq16}) for the model with 
competition, and for the model without competition
($T_{\rm linear}$). The number of votes (or money) lost by \
competition can be evaluated by looking at the difference between
$T$ and $T_{\rm linear}$. We see that there is a maximum loss when
$\Delta m = M/n$.
%

\subsection{Calculation of the expected number of votes $v$}
By integrating Eq.~(1) from the main text and performing a change of 
variables, we find that $v_i$ can be written as a function of $m_i$ as
\begin{equation}
v_i=v_i(0)+\frac{m_i}{\Delta m} - \frac{1}{\Delta m}\int_0^{m_i} \frac{S(m')}{n} dm'.
\label{eq17}
\end{equation}
Using Eq.~(\ref{eq12}), we can rewrite the above equation as
\begin{equation}
v_i=v_i(0)+\frac{1}{\Delta m} \left(1-\frac{S(0)}{n}\right) \int_0^{m_i} \exp\left[-\frac{1}{ n \Delta m } \int_0^{m'} r(m'')dm'' \right] dm'. 
\label{eq17}
\end{equation}
To find an analytical expression for $v$, we first decompose the external integral as
\begin{equation}
\begin{split}
v_i&=v_i(0)+ \frac{1}{\Delta m}\left(1-\frac{S(0)}{n}\right) \int_0^{m_{i-1}} \exp\left[-\frac{1}{ n \Delta m } \int_0^{m'} r(m'')dm'' \right] dm'\\ 
&+ \frac{1}{\Delta m}\left(1-\frac{S(0)}{n}\right) \int_{m_{i-1}}^{m_{i}} \exp\left[-\frac{1}{ n \Delta m } \int_0^{m'} r(m'')dm'' \right] dm',\\ 
\end{split}
\label{eq18}
\end{equation}
that compared with Eq.~(\ref{eq17}) can be rewritten as
\begin{equation}
\begin{split}
&v_i=v_i(0)-v_{i-1}(0)+v_{i-1}\\
&+ \frac{1}{\Delta m}\left(1-\frac{S(0)}{n}\right) \int_{m_{i-1}}^{m_{i}} \exp\left[-\frac{1}{ n \Delta m } \int_0^{m'} r(m'')dm'' \right] dm'.\\
\end{split}
\label{eq19}
\end{equation}
The result of this integral relies on the limits of the external 
integral. Using the definition of $r(m)$ for the external interval $m'\in[m_{i-1},m_{i}]$, we find that
\begin{equation}
\int_0^{m'} r(m'')dm'' = m_0+m_1+m_2+...+m_{i-1} + (N_c-i)m'.
\label{eq20}
\end{equation}
By solving the integrals, we finally find that the number of votes $v_i$ is
 given by
\begin{equation}
\begin{split}
&v_i=v_i(0)-v_{i-1}(0)+v_{i-1}\\
&-\frac{n-S(0)}{N_c-i} e^{-\sum_{j=0}^{i-1}m_j/( n \Delta m )} \left[e^{-\frac{(N_c-i)m_i}{ n \Delta m }}-e^{-\frac{(N_c-i)m_{i-1}}{ n \Delta m }}\right].
\end{split}
\label{eq21}
\end{equation}
As we can see from Eq.~(\ref{eq21}), the number of votes $v_i$ 
of a candidate $i$ is not only a function of his budget $m_i$, but 
also depends on the whole distribution $P(m)$. In
Fig.~\ref{figPol_5}B we show how $v(m)$ changes with $\Delta m$. 
As $\Delta m $ decreases, a large fraction of the voters become 
decided (i.e., $T\to 1$), and $v(m)$ displays a saturation for 
larges values of $m$ resulting on the diseconomy of scale due to the 
competition between candidates.

\section{Statistical comparison of models}
In order to compare our model with the simple case without  
competition, we make use of the Akaike's Information Criterion 
(AIC)~\cite{S_Motulsky2004}. The AIC is a model selection method that 
uses information theory to compare the relative estimation of the 
information lost by mathematical models used to generate data. 
Here, we used AIC to measure the relative quality of our model when 
compared with the linear non-competitive model.
Suppose that we have a model with $P$ parameters that fits a data 
set with $N$ points. Then, the AIC is defined as
\begin{equation}
AIC = N \ln\left(\frac{RSS}{N}\right) + 2 (P+1),
\label{eq22}
\end{equation}
where RSS is the \emph{residual sum of squares} given by
\begin{equation}
RSS = \sum_{i=1}^{N}(x_i - X_i)^2.
\end{equation}
Here, $x_i$ is the $i^{\rm th}$ value of the variable to be 
predicted and the $X_i$ is the predicted value of $x_i$. We  
calculate the AIC for each model using Eq.~(\ref{eq22}). Then, by 
Akaike's criterion, the preferred model is the one with the minimum 
AIC value. Here, we label the model without competition as WOC and 
the more complex model, where there is competition, as WC. The 
difference in AIC is then defined as
$\Delta AIC = AIC_{\rm WC} -  AIC_{\rm WOC}$. Once this difference 
is computed we calculate the probability that model WC minimize the 
information loss: 
\begin{equation}
P_{\rm WC} = \frac{e^{-0.5 \Delta AIC}}{1-e^{-0.5 \Delta AIC}}.
\end{equation}
Therefore, the probability that model WOC minimizes the information 
loss is $P_{\rm WOC}=1-P_{\rm WC}$. Here, we define the ratio 
between $P_{\rm WC}$ and $P_{\rm WOC}$ as the \emph{evidence ratio}, 
which means how many times the model WC is more likely to minimize 
the information loss. We then performed this analysis for federal 
and  state deputies for the 2014 elections in all 26 Brazilian 
states. The model WC and the model WOC are compared to the logarithm 
of the data (Tables~\ref{AIC_Fed_Dep_log}
and~\ref{AIC_Est_Dep_log}), and to the data without applying the 
logarithm (Tables~\ref{AIC_Fed_Dep}, and~\ref{AIC_Est_Dep}). The AIC 
shows that the model with competition best explains the data when 
compared to the linear model in all studied cases.

\section{Simulation on a Complex Network}
\begin{figure*}[!h]
\begin{center}
\includegraphics*[width=1\columnwidth]{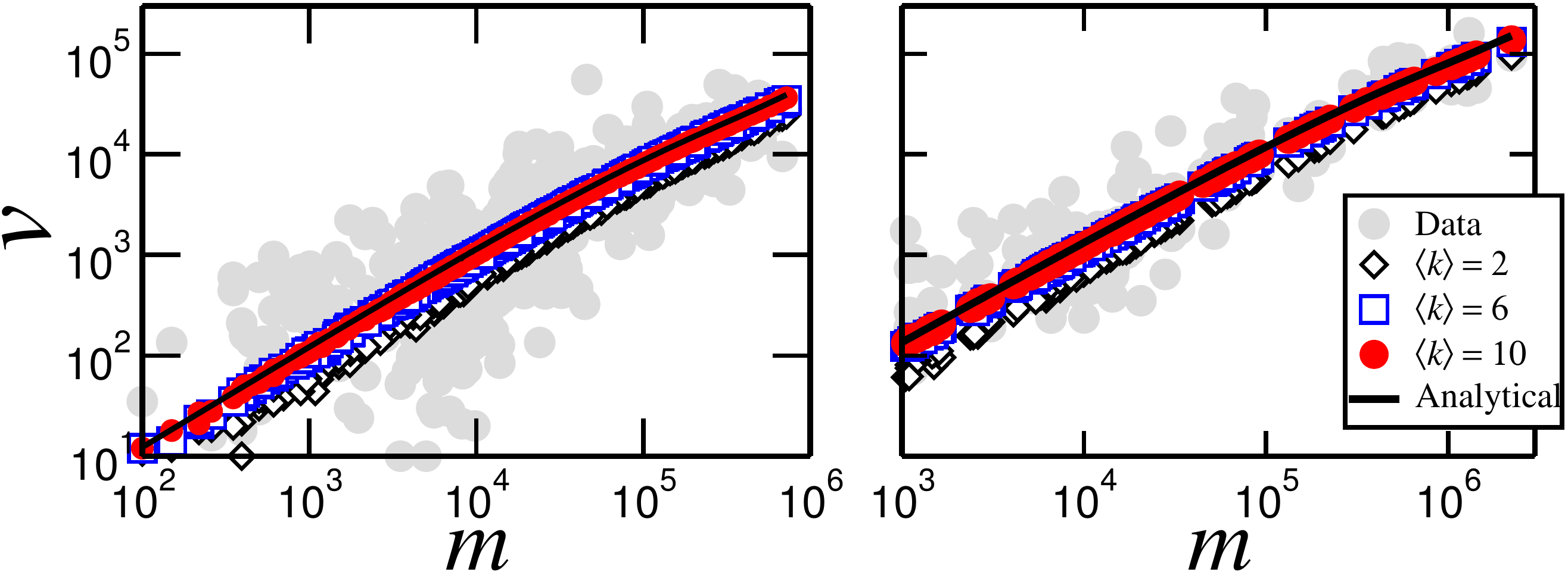}
\caption{\textbf{Simulation on a random network model.} Here we 
compare the analytical solution (black line) with the simulation on 
a random Erd\"os--R\'enyi network for the 2014 Esp\'irito Santo 
state election of federal deputies (A) and state deputies (B). Here, 
each gray circle represents the data for one candidate. We used 
three  different values of average connectivity:
$\langle k\rangle=2$ (black diamonds), $\langle k\rangle=6$ (blue 
squares) and $\langle k\rangle=10$ (red circles). Each symbol is the
result of a logarithmic binning for the money ($m$) axis over the 
simulation. We see that as we increase the average network degree, 
the simulation presents better agreement with the analytical 
solution. However, the analytical solution seems to capture the 
overall behavior for all networks tested. The apparent disagreement 
for $\langle k\rangle= 2$ is a consequence of a smaller effective 
size of the network, since an important fraction of nodes are not 
connected with the largest cluster. } 
\label{figS3}
\end{center}
\end{figure*}

\begin{figure*}[!h]
\begin{center}
\includegraphics*[width=1\columnwidth]{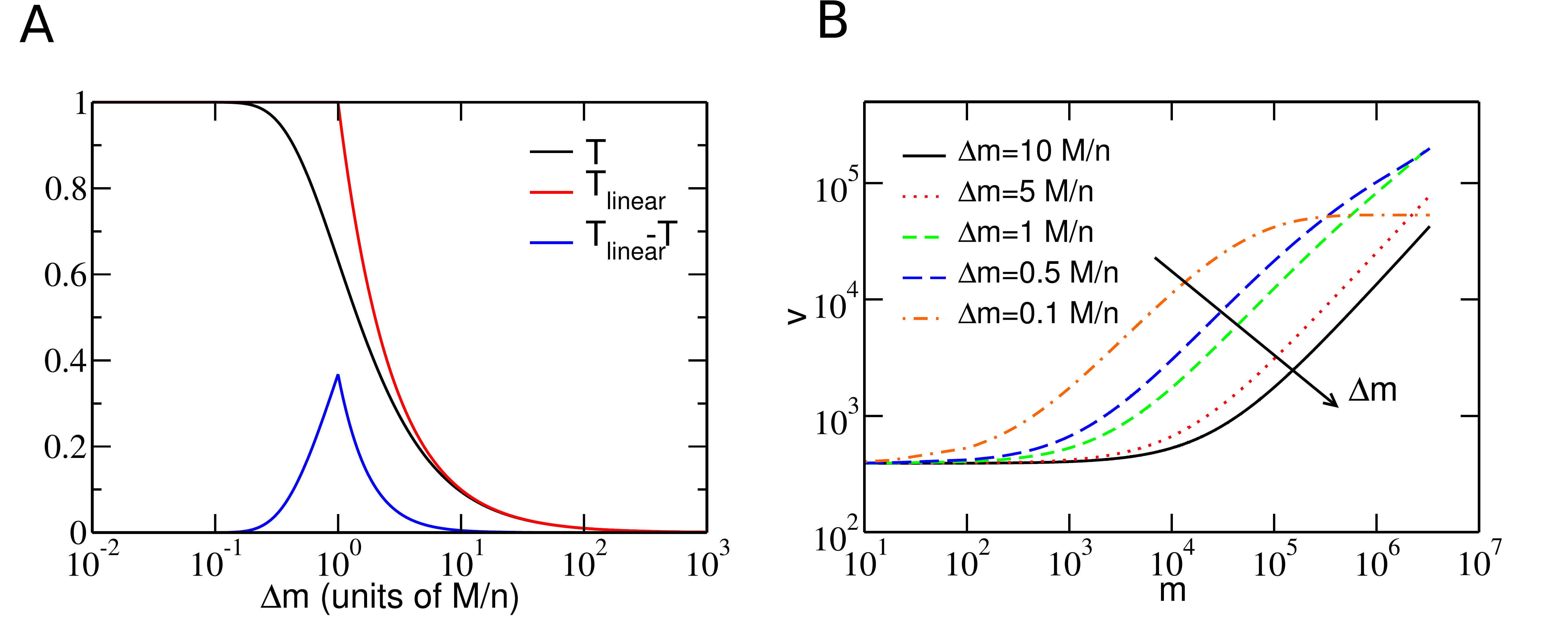}
\caption{\textbf{Dependence with $\Delta m$.} The solution of the 
mean field model enables us to calculate the turnout radio $T$ as a 
function of the dimensionless $n \Delta m /M$ parameter. In (A) we 
compare the turnout for the linear case where we excluded the 
competition between the candidates, $T_{linear}$, with the case with 
competition, $T$. The competition creates an exponential saturation, 
which increases the waste of money when candidates seek new voters. 
By looking at the difference $T_{linear} - T$, we can see that this 
inefficiency is maximum when $n \Delta m/M =1.0$. In (B) we show 
that as we decrease $ \Delta m $ the values of $v(m)$ usually 
increases, as expected by the definition of $ \Delta m$. However, 
there is a point where a saturation appears as the total number of 
votes starts to get close to the size of the system, resulting on 
a diseconomy of scale due to the competition between candidates.}
 \label{figPol_5}
\end{center}
\end{figure*}

In order to solve analytically the model, we make use of a mean 
field approximation where the network is a fully connected graph. To 
see if our solution still holds for a more complex topology, we 
performed simulations using the Erd\"os--R\'enyi network model with 
three different values for the average degree: $\langle k\rangle=2$, 
$6$ and $10$. As we can see in Fig.~\ref{figS3}A and B, for federal 
and state deputies, respectively, we find a good  agreement between 
the analytical solution (black line) and the real data (grey 
circles) for $\langle k\rangle=6$ and $10$. Due to computational 
performance, we chose the state of Esp\'irito Santo to perform the 
simulations. First, we made use of the candidates' budget for the 
2014 election as an input for the distribution of money $P(m)$. The 
network size is taken from the number of registered voters in 
Esp\'irito Santo, $N=2653536$, as presented in Table 1 and 2. Each 
candidate starts the simulation with only one node as a decided 
voter. This node is the initial seed for the candidate's marketing 
campaing.  The overall underestimation of the number of votes for 
$\langle k\rangle=2$ can be understood by noting that an important 
fraction of the network is made of unconnected nodes, therefore, for 
the candidates with seeds in the largest cluster the network seems 
to be smaller.

\section{Frequency distribution of votes}
\begin{figure*}[!h]
\begin{center}
\includegraphics*[width=1\columnwidth]{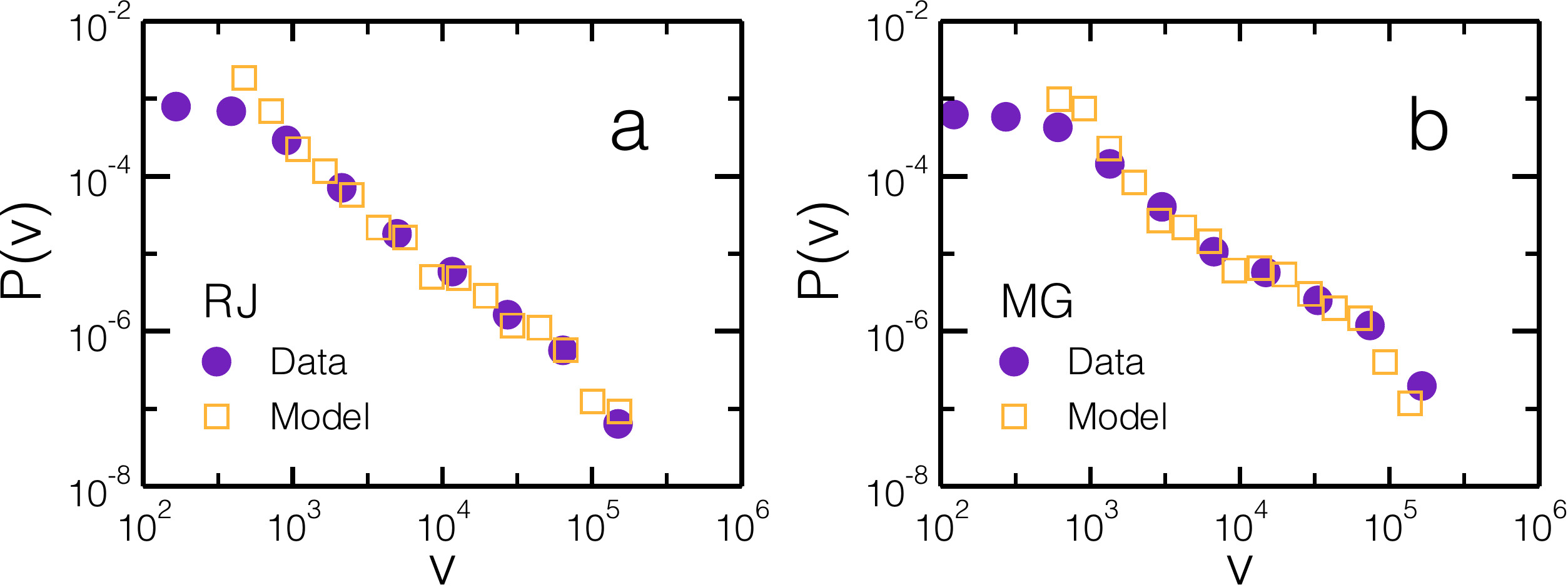}
\caption{{\bf Comparison between the actual distribution 
of votes with the ones obtained by our model.} Here, we 
show the comparison for the states of {\bf (a)} Rio de Janeiro and 
{\bf (b)} Minas Gerais. Again, the good agreement indicates that the 
long tail of $P(v)$ is a direct consequence of the money as an input 
for the dynamical process.}
 \label{figPol_6}
\end{center}
\end{figure*}

Here, we show the comparison between the empirical votes 
distribution for the states of Rio de Janeiro (Fig.~\ref{figPol_6}a) 
and Minas Gerais (Fig.~\ref{figPol_6}B) with the one obtained by 
our model. Again, the model reproduces correctly the empirical 
distribution of votes among candidates, $P(v)$.
%
\section{Study of the dispersion}
\begin{figure*}[!h]
\begin{center}
\includegraphics*[width=1\columnwidth]{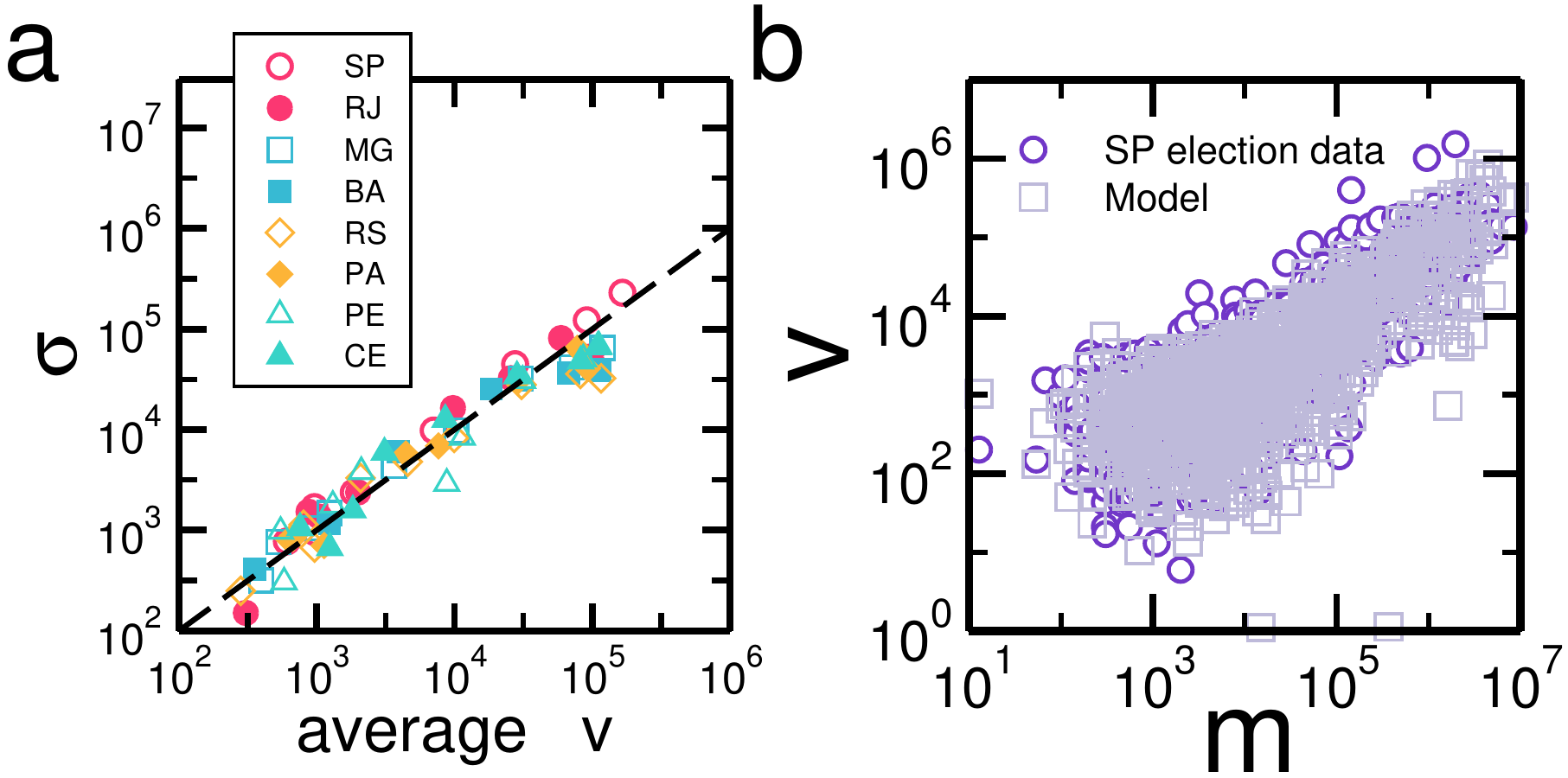}
\caption{{\bf Test of statistical dispersion.} It is widely known 
that the exponential distribution have the property that its mean 
and standard deviation are equal. Therefore we use this property in 
order to test if the dispersion along the mean follows an 
exponential distribution, as predicted by the MaxEnt hypothesis. In 
{\bf (a)} we see that for state deputies of the eight largest states 
in 2014 election the data is in close agreement with 
$\sigma=\langle v\rangle$ (dashed line). {\bf (b)} of votes 
calculate by our model to generate a random election. Here we show 
for the state of S\~ao Paulo that when we add random noise to our 
model (squares), we obtain a cloud that closely resembles the actual 
data (circles). }
 \label{figPol_7}
\end{center}
\end{figure*}

Our model allow us to calculate the mean or expected value of the 
number of votes. However, to fully describe the election we have 
also to study the statistical dispersion, which is given by the 
conditional probability distribution $p(v|m)$. We can use the 
concept of maximum entropy probability distribution (MaxEnt) from 
information theory to guess which is the $p(v|m)$ that maximizes the 
Shannon's Entropy~\cite{S_Jaynes1957}. Imposing only a constraint for 
the mean $\langle v\rangle $, the maximum entropy continuous 
distribution is exponential,
\begin{equation}
p(v|m)=\frac{1}{\langle v\rangle }e^{-\frac{v}{\langle v\rangle}},
\label{exp}
\end{equation}
which has the property that the mean and standard deviation are the 
same. We see in Figure~\ref{figPol_7}A that our data show a close 
linear relationship with approximately unit slope
$\sigma\approx \langle v\rangle  $, which strongly indicates that 
the Eq.~(\ref{exp}) accounts for all the random variation on $v(m)$ 
with the expected value calculated by our model. In the inset of 
Fig.~4F from the main text, we show these two elements in a simulation for the 
election of state deputy for the state of S\~ao Paulo, the greatest 
electoral college in Brazil. Figure~\ref{figPol_7}B shows that the 
addition of random dispersion to our model leads to a remarkable 
resemblance with real election data.

\newpage

\begin{table*}[]
\begin{tabular}{lllllll}

\cline{2-7}
                            & \multicolumn{1}{c}{}          & Federal deputies                  &                                 &                                     \\ \hline
\multicolumn{1}{|l|}{State} & \multicolumn{1}{l|}{n}        & \multicolumn{1}{l|}{M (R\$)}      & \multicolumn{1}{l|}{$S_f$}       & \multicolumn{1}{l|}{$T$ (\%)}       & \multicolumn{1}{l|}{r--pearson}       & \multicolumn{1}{l|}{p--value}      \\ \hline
\multicolumn{1}{|l|}{AC}    & \multicolumn{1}{l|}{506724}   & \multicolumn{1}{l|}{8480357.97}  & \multicolumn{1}{l|}{368332}   & \multicolumn{1}{l|}{72.6888799425}       & \multicolumn{1}{l|}{0.722748043319}       & \multicolumn{1}{l|}{3.30192976139e-11} \\ \hline
\multicolumn{1}{|l|}{AL}    & \multicolumn{1}{l|}{1995727}   & \multicolumn{1}{l|}{18421969.9}  & \multicolumn{1}{l|}{1283120}   & \multicolumn{1}{l|}{64.2933627696}       & \multicolumn{1}{l|}{0.840392141284}       & \multicolumn{1}{l|}{4.92302048603e-31} \\ \hline
\multicolumn{1}{|l|}{AM}    & \multicolumn{1}{l|}{2226891}   & \multicolumn{1}{l|}{23414726.56}  & \multicolumn{1}{l|}{1560085}   & \multicolumn{1}{l|}{70.0566395032}       & \multicolumn{1}{l|}{0.905459637663}       & \multicolumn{1}{l|}{9.23656309525e-31} \\ \hline
\multicolumn{1}{|l|}{AP}    & \multicolumn{1}{l|}{455514}   & \multicolumn{1}{l|}{8484530.19}  & \multicolumn{1}{l|}{368061}   & \multicolumn{1}{l|}{80.8012486993}       & \multicolumn{1}{l|}{0.566853357973}       & \multicolumn{1}{l|}{1.31431785604e-10} \\ \hline
\multicolumn{1}{|l|}{BA}    & \multicolumn{1}{l|}{10185417}   & \multicolumn{1}{l|}{72471496.94}  & \multicolumn{1}{l|}{5982371}   & \multicolumn{1}{l|}{58.7346693807}       & \multicolumn{1}{l|}{0.698084305668}       & \multicolumn{1}{l|}{2.28399775277e-49} \\ \hline
\multicolumn{1}{|l|}{CE}    & \multicolumn{1}{l|}{6271554}   & \multicolumn{1}{l|}{34838910.83}  & \multicolumn{1}{l|}{4002492}   & \multicolumn{1}{l|}{63.819780552}       & \multicolumn{1}{l|}{0.737909089987}       & \multicolumn{1}{l|}{7.95861881821e-36} \\ \hline
\multicolumn{1}{|l|}{ES}    & \multicolumn{1}{l|}{2653536}   & \multicolumn{1}{l|}{19490814.39}  & \multicolumn{1}{l|}{1665277}   & \multicolumn{1}{l|}{62.7569024879}       & \multicolumn{1}{l|}{0.822155443552}       & \multicolumn{1}{l|}{1.7418221573e-41} \\ \hline
\multicolumn{1}{|l|}{GO}    & \multicolumn{1}{l|}{4331733}   & \multicolumn{1}{l|}{65145051.12}  & \multicolumn{1}{l|}{2824329}   & \multicolumn{1}{l|}{65.2009022717}       & \multicolumn{1}{l|}{0.66788295422}       & \multicolumn{1}{l|}{6.05466767683e-20} \\ \hline
\multicolumn{1}{|l|}{MA}    & \multicolumn{1}{l|}{4497336}   & \multicolumn{1}{l|}{21197635.67}  & \multicolumn{1}{l|}{2836788}   & \multicolumn{1}{l|}{63.0770749617}       & \multicolumn{1}{l|}{0.685587130132}       & \multicolumn{1}{l|}{8.36126780457e-35} \\ \hline
\multicolumn{1}{|l|}{MG}    & \multicolumn{1}{l|}{15248681}   & \multicolumn{1}{l|}{160498695.1}  & \multicolumn{1}{l|}{9273472}   & \multicolumn{1}{l|}{60.8149124505}       & \multicolumn{1}{l|}{0.806652645383}       & \multicolumn{1}{l|}{1.27631105412e-147} \\ \hline
\multicolumn{1}{|l|}{MS}    & \multicolumn{1}{l|}{1818937}   & \multicolumn{1}{l|}{29384486.15}  & \multicolumn{1}{l|}{1174221}   & \multicolumn{1}{l|}{64.5553419387}       & \multicolumn{1}{l|}{0.778880439738}       & \multicolumn{1}{l|}{1.8509184753e-25} \\ \hline
\multicolumn{1}{|l|}{MT}    & \multicolumn{1}{l|}{2189703}   & \multicolumn{1}{l|}{27179850.24}  & \multicolumn{1}{l|}{1334861}   & \multicolumn{1}{l|}{60.9608243675}       & \multicolumn{1}{l|}{0.86687644675}       & \multicolumn{1}{l|}{1.61425424251e-33} \\ \hline
\multicolumn{1}{|l|}{PA}    & \multicolumn{1}{l|}{5188450}   & \multicolumn{1}{l|}{19219663.68}  & \multicolumn{1}{l|}{3496764}   & \multicolumn{1}{l|}{67.3951565496}       & \multicolumn{1}{l|}{0.714596611048}       & \multicolumn{1}{l|}{4.68308734801e-30} \\ \hline
\multicolumn{1}{|l|}{PB}    & \multicolumn{1}{l|}{2835882}   & \multicolumn{1}{l|}{14092397.88}  & \multicolumn{1}{l|}{1773112}   & \multicolumn{1}{l|}{62.5241811895}       & \multicolumn{1}{l|}{0.855261326688}       & \multicolumn{1}{l|}{2.61029697443e-30} \\ \hline
\multicolumn{1}{|l|}{PE}    & \multicolumn{1}{l|}{6356307}   & \multicolumn{1}{l|}{51507676.68}  & \multicolumn{1}{l|}{4129147}   & \multicolumn{1}{l|}{64.9614154886}       & \multicolumn{1}{l|}{0.728324391535}       & \multicolumn{1}{l|}{1.46267253887e-27} \\ \hline
\multicolumn{1}{|l|}{PI}    & \multicolumn{1}{l|}{2345694}   & \multicolumn{1}{l|}{24898627.07}  & \multicolumn{1}{l|}{1587477}   & \multicolumn{1}{l|}{67.6762186372}       & \multicolumn{1}{l|}{0.656433873665}       & \multicolumn{1}{l|}{1.22040618126e-13} \\ \hline
\multicolumn{1}{|l|}{PR}    & \multicolumn{1}{l|}{7865950}   & \multicolumn{1}{l|}{69592048.16}  & \multicolumn{1}{l|}{5275880}   & \multicolumn{1}{l|}{67.0723815941}       & \multicolumn{1}{l|}{0.728660777076}       & \multicolumn{1}{l|}{6.64008626188e-52} \\ \hline
\multicolumn{1}{|l|}{RJ}    & \multicolumn{1}{l|}{12141145}   & \multicolumn{1}{l|}{110784215.29}  & \multicolumn{1}{l|}{7063961}   & \multicolumn{1}{l|}{58.1820001326}       & \multicolumn{1}{l|}{0.56473987424}       & \multicolumn{1}{l|}{2.1291171572e-85} \\ \hline
\multicolumn{1}{|l|}{RN}    & \multicolumn{1}{l|}{2327451}   & \multicolumn{1}{l|}{14178893.28}  & \multicolumn{1}{l|}{1451341}   & \multicolumn{1}{l|}{62.3575319094}       & \multicolumn{1}{l|}{0.882530044098}       & \multicolumn{1}{l|}{1.40216619215e-30} \\ \hline
\multicolumn{1}{|l|}{RO}    & \multicolumn{1}{l|}{1127154}   & \multicolumn{1}{l|}{16967025.91}  & \multicolumn{1}{l|}{740924}   & \multicolumn{1}{l|}{65.7340523123}       & \multicolumn{1}{l|}{0.683327785935}       & \multicolumn{1}{l|}{7.96683605323e-13} \\ \hline
\multicolumn{1}{|l|}{RR}    & \multicolumn{1}{l|}{299558}   & \multicolumn{1}{l|}{8358613.48}  & \multicolumn{1}{l|}{225631}   & \multicolumn{1}{l|}{75.3213067252}       & \multicolumn{1}{l|}{0.598924952323}       & \multicolumn{1}{l|}{3.49851465097e-09} \\ \hline
\multicolumn{1}{|l|}{RS}    & \multicolumn{1}{l|}{8392033}   & \multicolumn{1}{l|}{57254432.25}  & \multicolumn{1}{l|}{5501353}   & \multicolumn{1}{l|}{65.554472915}       & \multicolumn{1}{l|}{0.836559267138}       & \multicolumn{1}{l|}{4.74667303986e-84} \\ \hline
\multicolumn{1}{|l|}{SC}    & \multicolumn{1}{l|}{4859324}   & \multicolumn{1}{l|}{31716424.53}  & \multicolumn{1}{l|}{3120297}   & \multicolumn{1}{l|}{64.2125736008}       & \multicolumn{1}{l|}{0.869812045153}       & \multicolumn{1}{l|}{2.11886214421e-41} \\ \hline
\multicolumn{1}{|l|}{SE}    & \multicolumn{1}{l|}{1454165}   & \multicolumn{1}{l|}{8057895.72}  & \multicolumn{1}{l|}{974311}   & \multicolumn{1}{l|}{67.0014063053}       & \multicolumn{1}{l|}{0.684912931565}       & \multicolumn{1}{l|}{2.44474497848e-12} \\ \hline
\multicolumn{1}{|l|}{TO}    & \multicolumn{1}{l|}{996887}   & \multicolumn{1}{l|}{15619685.1}  & \multicolumn{1}{l|}{670894}   & \multicolumn{1}{l|}{67.2989014803}       & \multicolumn{1}{l|}{0.76979251044}       & \multicolumn{1}{l|}{1.60978392322e-10} \\ \hline
\multicolumn{1}{|l|}{SP}    & \multicolumn{1}{l|}{31998432}   & \multicolumn{1}{l|}{241919492.64}  & \multicolumn{1}{l|}{19072393}   & \multicolumn{1}{l|}{59.6041487283}       & \multicolumn{1}{l|}{0.483246969693}       & \multicolumn{1}{l|}{9.81028340891e-81} \\ \hline
\end{tabular}
\centering
\caption{\textbf{Data description for Federal deputies.}
Here we describe the main properties of the data for the federal 
deputies election from all Brazilian states. For each state  we show 
the number of voters registered $n$, the total cost of the campaign 
in Brazilian Reais (R\$) $M$, the number of valid votes $S_f$, and 
the turnout percentage $T$.}
\label{States_table_Dep_fed}
\end{table*}

\begin{table*}[]
\begin{tabular}{lllllll}
\cline{2-7}
                            & \multicolumn{1}{c}{}          & State deputies                   &                                 &                                     \\ \hline
\multicolumn{1}{|l|}{State} & \multicolumn{1}{l|}{n}        & \multicolumn{1}{l|}{M (R\$)}      & \multicolumn{1}{l|}{$S_f$}       & \multicolumn{1}{l|}{$T$ (\%)}       & \multicolumn{1}{l|}{r--pearson}       & \multicolumn{1}{l|}{p--value}      \\ \hline
\multicolumn{1}{|l|}{AC}    & \multicolumn{1}{l|}{506724}   & \multicolumn{1}{l|}{10656037.7}  & \multicolumn{1}{l|}{377299}   & \multicolumn{1}{l|}{74.4584823296}       & \multicolumn{1}{l|}{0.803577951964}       & \multicolumn{1}{l|}{1.59880493578e-114} \\ \hline
\multicolumn{1}{|l|}{AL}    & \multicolumn{1}{l|}{1995727}   & \multicolumn{1}{l|}{19627276.99}  & \multicolumn{1}{l|}{1314659}   & \multicolumn{1}{l|}{65.8736891368}       & \multicolumn{1}{l|}{0.836240190525}       & \multicolumn{1}{l|}{8.11032735735e-76} \\ \hline
\multicolumn{1}{|l|}{AM}    & \multicolumn{1}{l|}{2226891}   & \multicolumn{1}{l|}{28001756.68}  & \multicolumn{1}{l|}{1547128}   & \multicolumn{1}{l|}{69.4747969254}       & \multicolumn{1}{l|}{0.498718455137}       & \multicolumn{1}{l|}{4.12728400638e-39} \\ \hline
\multicolumn{1}{|l|}{AP}    & \multicolumn{1}{l|}{455514}   & \multicolumn{1}{l|}{5626676.58}  & \multicolumn{1}{l|}{373731}   & \multicolumn{1}{l|}{82.0459963909}       & \multicolumn{1}{l|}{0.647712202199}       & \multicolumn{1}{l|}{7.66012489633e-44} \\ \hline
\multicolumn{1}{|l|}{BA}    & \multicolumn{1}{l|}{10185417}   & \multicolumn{1}{l|}{47294333.36}  & \multicolumn{1}{l|}{6053428}   & \multicolumn{1}{l|}{59.432304048}       & \multicolumn{1}{l|}{0.782568469296}       & \multicolumn{1}{l|}{1.25445144149e-126} \\ \hline
\multicolumn{1}{|l|}{CE}    & \multicolumn{1}{l|}{6271554}   & \multicolumn{1}{l|}{32576249.09}  & \multicolumn{1}{l|}{4095292}   & \multicolumn{1}{l|}{65.2994776095}       & \multicolumn{1}{l|}{0.686934485759}       & \multicolumn{1}{l|}{7.21069456863e-83} \\ \hline
\multicolumn{1}{|l|}{ES}    & \multicolumn{1}{l|}{2653536}   & \multicolumn{1}{l|}{23289124.65}  & \multicolumn{1}{l|}{1748232}   & \multicolumn{1}{l|}{65.8831084259}       & \multicolumn{1}{l|}{0.741623876986}       & \multicolumn{1}{l|}{4.30172166816e-85} \\ \hline
\multicolumn{1}{|l|}{GO}    & \multicolumn{1}{l|}{4331733}   & \multicolumn{1}{l|}{79310623.34}  & \multicolumn{1}{l|}{2882804}   & \multicolumn{1}{l|}{66.550823885}       & \multicolumn{1}{l|}{0.734121134203}       & \multicolumn{1}{l|}{3.1534575998e-129} \\ \hline
\multicolumn{1}{|l|}{MA}    & \multicolumn{1}{l|}{4497336}   & \multicolumn{1}{l|}{25979148.94}  & \multicolumn{1}{l|}{2917772}   & \multicolumn{1}{l|}{64.8777854268}       & \multicolumn{1}{l|}{0.839414632196}       & \multicolumn{1}{l|}{1.67610556861e-134} \\ \hline
\multicolumn{1}{|l|}{MG}    & \multicolumn{1}{l|}{15248681}   & \multicolumn{1}{l|}{177676580.98}  & \multicolumn{1}{l|}{9283721}   & \multicolumn{1}{l|}{60.8821248212}       & \multicolumn{1}{l|}{0.224029765254}       & \multicolumn{1}{l|}{4.52706918339e-14} \\ \hline
\multicolumn{1}{|l|}{MS}    & \multicolumn{1}{l|}{1818937}   & \multicolumn{1}{l|}{45948066.57}  & \multicolumn{1}{l|}{1204007}   & \multicolumn{1}{l|}{66.1928917824}       & \multicolumn{1}{l|}{0.799004169269}       & \multicolumn{1}{l|}{6.08398880781e-90} \\ \hline
\multicolumn{1}{|l|}{MT}    & \multicolumn{1}{l|}{2189703}   & \multicolumn{1}{l|}{51639423.61}  & \multicolumn{1}{l|}{1375357}   & \multicolumn{1}{l|}{62.8102075944}       & \multicolumn{1}{l|}{0.771404926583}       & \multicolumn{1}{l|}{2.02454597864e-62} \\ \hline
\multicolumn{1}{|l|}{PA}    & \multicolumn{1}{l|}{5188450}   & \multicolumn{1}{l|}{31595425.94}  & \multicolumn{1}{l|}{3453031}   & \multicolumn{1}{l|}{66.5522651274}       & \multicolumn{1}{l|}{0.715064099361}       & \multicolumn{1}{l|}{1.78371156902e-110} \\ \hline
\multicolumn{1}{|l|}{PB}    & \multicolumn{1}{l|}{2835882}   & \multicolumn{1}{l|}{17219860.72}  & \multicolumn{1}{l|}{1835376}   & \multicolumn{1}{l|}{64.7197591437}       & \multicolumn{1}{l|}{0.782772014141}       & \multicolumn{1}{l|}{1.67491078919e-74} \\ \hline
\multicolumn{1}{|l|}{PE}    & \multicolumn{1}{l|}{6356307}   & \multicolumn{1}{l|}{40641680.29}  & \multicolumn{1}{l|}{4171737}   & \multicolumn{1}{l|}{65.6314586441}       & \multicolumn{1}{l|}{0.748176330675}       & \multicolumn{1}{l|}{1.36075786856e-90} \\ \hline
\multicolumn{1}{|l|}{PI}    & \multicolumn{1}{l|}{2345694}   & \multicolumn{1}{l|}{20320016.99}  & \multicolumn{1}{l|}{1607165}   & \multicolumn{1}{l|}{68.5155438007}       & \multicolumn{1}{l|}{0.816180285209}       & \multicolumn{1}{l|}{2.19894569952e-58} \\ \hline
\multicolumn{1}{|l|}{PR}    & \multicolumn{1}{l|}{7865950}   & \multicolumn{1}{l|}{61749634.55}  & \multicolumn{1}{l|}{5298846}   & \multicolumn{1}{l|}{67.3643488708}       & \multicolumn{1}{l|}{0.878519254427}       & \multicolumn{1}{l|}{1.18750225791e-247} \\ \hline
\multicolumn{1}{|l|}{RJ}    & \multicolumn{1}{l|}{12141145}   & \multicolumn{1}{l|}{130048101.34}  & \multicolumn{1}{l|}{7122375}   & \multicolumn{1}{l|}{58.6631244417}       & \multicolumn{1}{l|}{0.572037787678}       & \multicolumn{1}{l|}{5.93043422869e-167} \\ \hline
\multicolumn{1}{|l|}{RN}    & \multicolumn{1}{l|}{2327451}   & \multicolumn{1}{l|}{18343797.5}  & \multicolumn{1}{l|}{1529149}   & \multicolumn{1}{l|}{65.700588326}       & \multicolumn{1}{l|}{0.850127492405}       & \multicolumn{1}{l|}{3.79357570719e-72} \\ \hline
\multicolumn{1}{|l|}{RO}    & \multicolumn{1}{l|}{1127154}   & \multicolumn{1}{l|}{25138956.64}  & \multicolumn{1}{l|}{761590}   & \multicolumn{1}{l|}{67.5675196113}       & \multicolumn{1}{l|}{0.741913371628}       & \multicolumn{1}{l|}{3.21262552227e-70} \\ \hline
\multicolumn{1}{|l|}{RR}    & \multicolumn{1}{l|}{299558}   & \multicolumn{1}{l|}{13376926.76}  & \multicolumn{1}{l|}{242398}   & \multicolumn{1}{l|}{80.9185533352}       & \multicolumn{1}{l|}{0.813681589893}       & \multicolumn{1}{l|}{8.41908443078e-96} \\ \hline
\multicolumn{1}{|l|}{RS}    & \multicolumn{1}{l|}{8392033}   & \multicolumn{1}{l|}{54552702.15}  & \multicolumn{1}{l|}{5592657}   & \multicolumn{1}{l|}{66.6424571972}       & \multicolumn{1}{l|}{0.691245148201}       & \multicolumn{1}{l|}{3.6535374402e-98} \\ \hline
\multicolumn{1}{|l|}{SC}    & \multicolumn{1}{l|}{4859324}   & \multicolumn{1}{l|}{52245781.28}  & \multicolumn{1}{l|}{3280653}   & \multicolumn{1}{l|}{67.5125387811}       & \multicolumn{1}{l|}{0.816495812077}       & \multicolumn{1}{l|}{1.75144104038e-102} \\ \hline
\multicolumn{1}{|l|}{SE}    & \multicolumn{1}{l|}{1454165}   & \multicolumn{1}{l|}{8833829.91}  & \multicolumn{1}{l|}{967550}   & \multicolumn{1}{l|}{66.5364659444}       & \multicolumn{1}{l|}{0.716103939585}       & \multicolumn{1}{l|}{1.17133851209e-28} \\ \hline
\multicolumn{1}{|l|}{TO}    & \multicolumn{1}{l|}{996887}   & \multicolumn{1}{l|}{20185053.82}  & \multicolumn{1}{l|}{699008}   & \multicolumn{1}{l|}{70.1190806982}       & \multicolumn{1}{l|}{0.864802148991}       & \multicolumn{1}{l|}{1.46231152842e-78} \\ \hline
\multicolumn{1}{|l|}{SP}    & \multicolumn{1}{l|}{31998432}   & \multicolumn{1}{l|}{231516634.41}  & \multicolumn{1}{l|}{17618073}   & \multicolumn{1}{l|}{55.0591760246}       & \multicolumn{1}{l|}{0.722245302764}       & \multicolumn{1}{l|}{1.00059011444e-314} \\ \hline
\end{tabular}
\centering
\caption{\textbf{Data description for State deputies.} Here we 
describe the main properties of the data for the state deputies 
election from all Brazilian states. For each state  we show the 
number of voters registered $n$, the total cost of the campaign in 
Brazilian Reais (R\$) $M$, the number of valid votes $S_f$, and the 
turnout percentage $T$.}
\label{States_table_Dep_est}
\end{table*}

\begin{table*}[]

\begin{tabular}{lllll}
\cline{2-5}
                         & \multicolumn{4}{c}{Federal deputies}                                                                                                                     \\ \hline
\multicolumn{1}{|l|}{State} & \multicolumn{1}{|l|}{$\Delta$ AIC} & \multicolumn{1}{|l|}{Probability WOC}              & \multicolumn{1}{|l|}{Probability WC}            & \multicolumn{1}{|l|}{Evidence radio}                         \\ \hline
\multicolumn{1}{|l|}{AC} & \multicolumn{1}{l|}{6.1827384262}  & \multicolumn{1}{l|}{0.0434646736}      & \multicolumn{1}{l|}{0.9565353264} & \multicolumn{1}{l|}{22.0071898888}     \\ \hline
\multicolumn{1}{|l|}{AL} & \multicolumn{1}{l|}{4.7608349884}  & \multicolumn{1}{l|}{0.0846782011}      & \multicolumn{1}{l|}{0.9153217989} & \multicolumn{1}{l|}{10.8094147898}     \\ \hline
\multicolumn{1}{|l|}{AM} & \multicolumn{1}{l|}{2.6647303199}  & \multicolumn{1}{l|}{0.2087684091}      & \multicolumn{1}{l|}{0.7912315909} & \multicolumn{1}{l|}{3.7899967439}      \\ \hline
\multicolumn{1}{|l|}{AP} & \multicolumn{1}{l|}{2.806065353}   & \multicolumn{1}{l|}{0.1973353125}      & \multicolumn{1}{l|}{0.8026646875} & \multicolumn{1}{l|}{4.0675167435}      \\ \hline
\multicolumn{1}{|l|}{CE} & \multicolumn{1}{l|}{10.5123485714} & \multicolumn{1}{l|}{0.0051881611}      & \multicolumn{1}{l|}{0.9948118389} & \multicolumn{1}{l|}{191.7465189}       \\ \hline
\multicolumn{1}{|l|}{ES} & \multicolumn{1}{l|}{14.4468382526} & \multicolumn{1}{l|}{0.0007287731}      & \multicolumn{1}{l|}{0.9992712269} & \multicolumn{1}{l|}{1371.1692559}      \\ \hline
\multicolumn{1}{|l|}{GO} & \multicolumn{1}{l|}{7.4677125018}  & \multicolumn{1}{l|}{0.0233425922}      & \multicolumn{1}{l|}{0.9766574078} & \multicolumn{1}{l|}{41.8401435584}     \\ \hline
\multicolumn{1}{|l|}{MA} & \multicolumn{1}{l|}{7.7592626578}  & \multicolumn{1}{l|}{0.0202403068}      & \multicolumn{1}{l|}{0.9797596932} & \multicolumn{1}{l|}{48.4063657539}     \\ \hline
\multicolumn{1}{|l|}{MS} & \multicolumn{1}{l|}{10.1109475602} & \multicolumn{1}{l|}{0.0063339711}      & \multicolumn{1}{l|}{0.9936660289} & \multicolumn{1}{l|}{156.878838813}     \\ \hline
\multicolumn{1}{|l|}{MT} & \multicolumn{1}{l|}{3.9037010608}  & \multicolumn{1}{l|}{0.1243517168}      & \multicolumn{1}{l|}{0.8756482832} & \multicolumn{1}{l|}{7.0417064229}      \\ \hline
\multicolumn{1}{|l|}{PA} & \multicolumn{1}{l|}{9.663695483}   & \multicolumn{1}{l|}{0.0079087312}      & \multicolumn{1}{l|}{0.9920912688} & \multicolumn{1}{l|}{125.442532017}     \\ \hline
\multicolumn{1}{|l|}{PB} & \multicolumn{1}{l|}{4.6657768971}  & \multicolumn{1}{l|}{0.0884355349}      & \multicolumn{1}{l|}{0.9115644651} & \multicolumn{1}{l|}{10.3076717549}     \\ \hline
\multicolumn{1}{|l|}{PI} & \multicolumn{1}{l|}{2.3415470513}  & \multicolumn{1}{l|}{0.2367151943}      & \multicolumn{1}{l|}{0.7632848057} & \multicolumn{1}{l|}{3.2244858966}      \\ \hline
\multicolumn{1}{|l|}{RN} & \multicolumn{1}{l|}{5.1455026334}  & \multicolumn{1}{l|}{0.0709128217}      & \multicolumn{1}{l|}{0.9290871783} & \multicolumn{1}{l|}{13.1018221599}     \\ \hline
\multicolumn{1}{|l|}{RO} & \multicolumn{1}{l|}{9.807097669}   & \multicolumn{1}{l|}{0.0073655493}      & \multicolumn{1}{l|}{0.9926344507} & \multicolumn{1}{l|}{134.767198525}     \\ \hline
\multicolumn{1}{|l|}{RR} & \multicolumn{1}{l|}{1.8879845033}  & \multicolumn{1}{l|}{0.2800946322}      & \multicolumn{1}{l|}{0.7199053678} & \multicolumn{1}{l|}{2.5702219362}      \\ \hline
\multicolumn{1}{|l|}{SC} & \multicolumn{1}{l|}{13.3469679287} & \multicolumn{1}{l|}{0.0012623917}      & \multicolumn{1}{l|}{0.9987376083} & \multicolumn{1}{l|}{791.147136976}     \\ \hline
\multicolumn{1}{|l|}{SE} & \multicolumn{1}{l|}{2.8635043257}  & \multicolumn{1}{l|}{0.1928258238}      & \multicolumn{1}{l|}{0.8071741762} & \multicolumn{1}{l|}{4.1860273715}      \\ \hline
\multicolumn{1}{|l|}{TO} & \multicolumn{1}{l|}{5.8040508651}  & \multicolumn{1}{l|}{0.0520535295}      & \multicolumn{1}{l|}{0.9479464705} & \multicolumn{1}{l|}{18.2109931789}     \\ \hline
\multicolumn{1}{|l|}{BA} & \multicolumn{1}{l|}{16.0404485774} & \multicolumn{1}{l|}{0.0003286382}      & \multicolumn{1}{l|}{0.9996713618} & \multicolumn{1}{l|}{3041.85951117}     \\ \hline
\multicolumn{1}{|l|}{MG} & \multicolumn{1}{l|}{32.2756994687} & \multicolumn{1}{l|}{9.80439653939e-08} & \multicolumn{1}{l|}{0.999999902}  & \multicolumn{1}{l|}{10199504.8644}     \\ \hline
\multicolumn{1}{|l|}{SP} & \multicolumn{1}{l|}{74.5043113536} & \multicolumn{1}{l|}{6.63123395728e-17} & \multicolumn{1}{l|}{1}            & \multicolumn{1}{l|}{1.50801495837e+16} \\ \hline
\multicolumn{1}{|l|}{RJ} & \multicolumn{1}{l|}{42.5533963117} & \multicolumn{1}{l|}{5.74972928891e-10} & \multicolumn{1}{l|}{0.9999999994} & \multicolumn{1}{l|}{1739212316.23}     \\ \hline
\multicolumn{1}{|l|}{RS} & \multicolumn{1}{l|}{18.8727293265} & \multicolumn{1}{l|}{7.97635097082e-05} & \multicolumn{1}{l|}{0.9999202365} & \multicolumn{1}{l|}{12536.0611657}     \\ \hline
\multicolumn{1}{|l|}{PE} & \multicolumn{1}{l|}{7.192496085}   & \multicolumn{1}{l|}{0.026694303}       & \multicolumn{1}{l|}{0.973305697}  & \multicolumn{1}{l|}{36.4611767018}     \\ \hline
\multicolumn{1}{|l|}{PR} & \multicolumn{1}{l|}{15.899933541}  & \multicolumn{1}{l|}{0.0003525495}      & \multicolumn{1}{l|}{0.9996474505} & \multicolumn{1}{l|}{2835.48072726}     \\ \hline
\end{tabular}
\centering
\caption{{\bf Statistical comparison between the models.} We use the Akaike's information criterion (AIC) to compare the two models: WOC (without competition) and WC (with competition). The AIC lets us determine which model is more likely to  describe correctly the data and quantify by calculating the probabilities and an evidence radio. The probability column shows the likelihood of each model to be the most correctly. The evidence radio is the fraction of Probability WC by Probability WOC, which means how many times model WC is likely to be correct than model WOC. Here, the AIC was applied in the logarithm of the data.  }
\label{AIC_Fed_Dep_log}
\end{table*}

\begin{table*}[]

\begin{tabular}{lllll}
\cline{2-5}
                         & \multicolumn{4}{c}{States deputies}                                                                                                                     \\ \hline
\multicolumn{1}{|l|}{State} & \multicolumn{1}{|l|}{$\Delta$ AIC} & \multicolumn{1}{|l|}{Probability WOC}              & \multicolumn{1}{|l|}{Probability WC}            & \multicolumn{1}{|l|}{Evidence radio}                         \\ \hline

\multicolumn{1}{|l|}{AC} & \multicolumn{1}{l|}{40.1333350824} & \multicolumn{1}{l|}{1.92822192099e-09} & \multicolumn{1}{l|}{0.9999999981} & \multicolumn{1}{l|}{518612503.667}     \\ \hline
\multicolumn{1}{|l|}{AL} & \multicolumn{1}{l|}{10.5644673228} & \multicolumn{1}{l|}{0.005055382}       & \multicolumn{1}{l|}{0.994944618}  & \multicolumn{1}{l|}{196.808989398}     \\ \hline
\multicolumn{1}{|l|}{AM} & \multicolumn{1}{l|}{25.7812709128} & \multicolumn{1}{l|}{2.52154694444e-06} & \multicolumn{1}{l|}{0.9999974785} & \multicolumn{1}{l|}{396580.948318}     \\ \hline
\multicolumn{1}{|l|}{AP} & \multicolumn{1}{l|}{13.4280658125} & \multicolumn{1}{l|}{0.0012122878}      & \multicolumn{1}{l|}{0.9987877122} & \multicolumn{1}{l|}{823.886605911}     \\ \hline
\multicolumn{1}{|l|}{CE} & \multicolumn{1}{l|}{24.7191563734} & \multicolumn{1}{l|}{4.28846166679e-06} & \multicolumn{1}{l|}{0.9999957115} & \multicolumn{1}{l|}{233182.849525}     \\ \hline
\multicolumn{1}{|l|}{ES} & \multicolumn{1}{l|}{40.567665947}  & \multicolumn{1}{l|}{1.55182686878e-09} & \multicolumn{1}{l|}{0.9999999984} & \multicolumn{1}{l|}{644401781.26}      \\ \hline
\multicolumn{1}{|l|}{GO} & \multicolumn{1}{l|}{22.588786324}  & \multicolumn{1}{l|}{1.24423372754e-05} & \multicolumn{1}{l|}{0.9999875577} & \multicolumn{1}{l|}{80369.751722}      \\ \hline
\multicolumn{1}{|l|}{MA} & \multicolumn{1}{l|}{17.3075446836} & \multicolumn{1}{l|}{0.000174437}       & \multicolumn{1}{l|}{0.999825563}  & \multicolumn{1}{l|}{5731.72803942}     \\ \hline
\multicolumn{1}{|l|}{MS} & \multicolumn{1}{l|}{29.9037339555} & \multicolumn{1}{l|}{3.20986330536e-07} & \multicolumn{1}{l|}{0.999999679}  & \multicolumn{1}{l|}{3115396.46359}     \\ \hline
\multicolumn{1}{|l|}{MT} & \multicolumn{1}{l|}{19.5744050013} & \multicolumn{1}{l|}{5.61626470736e-05} & \multicolumn{1}{l|}{0.9999438374} & \multicolumn{1}{l|}{17804.4285563}     \\ \hline
\multicolumn{1}{|l|}{PA} & \multicolumn{1}{l|}{31.4296195953} & \multicolumn{1}{l|}{1.49673445309e-07} & \multicolumn{1}{l|}{0.9999998503} & \multicolumn{1}{l|}{6681210.87384}     \\ \hline
\multicolumn{1}{|l|}{PB} & \multicolumn{1}{l|}{18.9409648861} & \multicolumn{1}{l|}{7.7088261887e-05}  & \multicolumn{1}{l|}{0.9999229117} & \multicolumn{1}{l|}{12971.1435601}     \\ \hline
\multicolumn{1}{|l|}{PI} & \multicolumn{1}{l|}{8.8288917589}  & \multicolumn{1}{l|}{0.0119565683}      & \multicolumn{1}{l|}{0.9880434317} & \multicolumn{1}{l|}{82.6360378881}     \\ \hline
\multicolumn{1}{|l|}{RN} & \multicolumn{1}{l|}{8.8191591598}  & \multicolumn{1}{l|}{0.0120141936}      & \multicolumn{1}{l|}{0.9879858064} & \multicolumn{1}{l|}{82.2348830361}     \\ \hline
\multicolumn{1}{|l|}{RO} & \multicolumn{1}{l|}{24.1600601847} & \multicolumn{1}{l|}{5.67162001552e-06} & \multicolumn{1}{l|}{0.9999943284} & \multicolumn{1}{l|}{176315.466418}     \\ \hline
\multicolumn{1}{|l|}{RR} & \multicolumn{1}{l|}{20.058166039}  & \multicolumn{1}{l|}{4.4096633465e-05}  & \multicolumn{1}{l|}{0.9999559034} & \multicolumn{1}{l|}{22676.468129}      \\ \hline
\multicolumn{1}{|l|}{SC} & \multicolumn{1}{l|}{30.7721830943} & \multicolumn{1}{l|}{2.07924302228e-07} & \multicolumn{1}{l|}{0.9999997921} & \multicolumn{1}{l|}{4809441.61582}     \\ \hline
\multicolumn{1}{|l|}{SE} & \multicolumn{1}{l|}{8.7956277831}  & \multicolumn{1}{l|}{0.0121546554}      & \multicolumn{1}{l|}{0.9878453446} & \multicolumn{1}{l|}{81.2730027198}     \\ \hline
\multicolumn{1}{|l|}{TO} & \multicolumn{1}{l|}{13.5362503146} & \multicolumn{1}{l|}{0.0011485278}      & \multicolumn{1}{l|}{0.9988514722} & \multicolumn{1}{l|}{869.679851625}     \\ \hline
\multicolumn{1}{|l|}{BA} & \multicolumn{1}{l|}{42.8429750252} & \multicolumn{1}{l|}{4.97469178945e-10} & \multicolumn{1}{l|}{0.9999999995} & \multicolumn{1}{l|}{2010174784.34}     \\ \hline
\multicolumn{1}{|l|}{MG} & \multicolumn{1}{l|}{55.735515674}  & \multicolumn{1}{l|}{7.89199040127e-13} & \multicolumn{1}{l|}{1}            & \multicolumn{1}{l|}{1.26710747119e+12} \\ \hline
\multicolumn{1}{|l|}{SP} & \multicolumn{1}{l|}{168.97837878}  & \multicolumn{1}{l|}{2.0268017352e-37}  & \multicolumn{1}{l|}{1}            & \multicolumn{1}{l|}{4.93388170453e+36} \\ \hline
\multicolumn{1}{|l|}{RJ} & \multicolumn{1}{l|}{82.3116713751} & \multicolumn{1}{l|}{1.33735794708e-18} & \multicolumn{1}{l|}{1}            & \multicolumn{1}{l|}{7.47742967531e+17} \\ \hline
\multicolumn{1}{|l|}{RS} & \multicolumn{1}{l|}{52.8467317602} & \multicolumn{1}{l|}{3.3456307332e-12}  & \multicolumn{1}{l|}{1}            & \multicolumn{1}{l|}{298897302106}      \\ \hline
\multicolumn{1}{|l|}{PE} & \multicolumn{1}{l|}{19.5248749346} & \multicolumn{1}{l|}{5.75708013522e-05} & \multicolumn{1}{l|}{0.9999424292} & \multicolumn{1}{l|}{17368.9162859}     \\ \hline
\multicolumn{1}{|l|}{PR} & \multicolumn{1}{l|}{42.819329153}  & \multicolumn{1}{l|}{5.0338563126e-10}  & \multicolumn{1}{l|}{0.9999999995} & \multicolumn{1}{l|}{1986548557.2}      \\ \hline
\end{tabular}
\centering
\caption{Statistical comparison between the models. We used the Akaike's information criterion (AIC) to compare the two models: WOC (without competition) and WC (with competition). The AIC lets us determine which model is more likely to  describe correctly the data and quantify by calculating the probabilities and an evidence radio. The probability column shows the likelihood of each model to be the most correctly. The evidence radio is the fraction of Probability WC by Probability WOC, which means how many times model WC is likely to be correct than model WOC. Here, the AIC was applied in the logarithm of the data.  }
\label{AIC_Est_Dep_log}
\end{table*}

\begin{table*}[]

\begin{tabular}{lllll}
\cline{2-5}
                         & \multicolumn{4}{c}{Federal deputies}                                                                                                                     \\ \hline
\multicolumn{1}{|l|}{State} & \multicolumn{1}{|l|}{$\Delta$ AIC} & \multicolumn{1}{|l|}{Probability WOC}              & \multicolumn{1}{|l|}{Probability WC}            & \multicolumn{1}{|l|}{Evidence radio}                         \\ \hline
\multicolumn{1}{|l|}{AC} & \multicolumn{1}{l|}{50.1494383861} & \multicolumn{1}{l|}{1.28880680099e-11}  & \multicolumn{1}{l|}{1} & \multicolumn{1}{l|}{77591148589.4}      \\ \hline
\multicolumn{1}{|l|}{AL} & \multicolumn{1}{l|}{101.196392792} & \multicolumn{1}{l|}{1.0604312338e-22}   & \multicolumn{1}{l|}{1} & \multicolumn{1}{l|}{9.43012585939e+21}  \\ \hline
\multicolumn{1}{|l|}{AM} & \multicolumn{1}{l|}{107.771152638} & \multicolumn{1}{l|}{3.96087877268e-24}  & \multicolumn{1}{l|}{1} & \multicolumn{1}{l|}{2.524692265e+23}    \\ \hline
\multicolumn{1}{|l|}{AP} & \multicolumn{1}{l|}{120.857209596} & \multicolumn{1}{l|}{5.70414277247e-27}  & \multicolumn{1}{l|}{1} & \multicolumn{1}{l|}{1.75311179942e+26}  \\ \hline
\multicolumn{1}{|l|}{CE} & \multicolumn{1}{l|}{184.905492905} & \multicolumn{1}{l|}{7.05151410541e-41}  & \multicolumn{1}{l|}{1} & \multicolumn{1}{l|}{1.41813514807e+40}  \\ \hline
\multicolumn{1}{|l|}{ES} & \multicolumn{1}{l|}{202.215834338} & \multicolumn{1}{l|}{1.22854055302e-44}  & \multicolumn{1}{l|}{1} & \multicolumn{1}{l|}{8.13973944563e+43}  \\ \hline
\multicolumn{1}{|l|}{GO} & \multicolumn{1}{l|}{79.8068548356} & \multicolumn{1}{l|}{4.67909285203e-18}  & \multicolumn{1}{l|}{1} & \multicolumn{1}{l|}{2.13716639448e+17}  \\ \hline
\multicolumn{1}{|l|}{MA} & \multicolumn{1}{l|}{224.617775801} & \multicolumn{1}{l|}{1.67830046988e-49}  & \multicolumn{1}{l|}{1} & \multicolumn{1}{l|}{5.95840862792e+48}  \\ \hline
\multicolumn{1}{|l|}{MS} & \multicolumn{1}{l|}{118.603128883} & \multicolumn{1}{l|}{1.76058823276e-26}  & \multicolumn{1}{l|}{1} & \multicolumn{1}{l|}{5.67991982108e+25}  \\ \hline
\multicolumn{1}{|l|}{MT} & \multicolumn{1}{l|}{120.825144982} & \multicolumn{1}{l|}{5.79633035462e-27}  & \multicolumn{1}{l|}{1} & \multicolumn{1}{l|}{1.72522947938e+26}  \\ \hline
\multicolumn{1}{|l|}{PA} & \multicolumn{1}{l|}{141.83521874}  & \multicolumn{1}{l|}{1.58808440446e-31}  & \multicolumn{1}{l|}{1} & \multicolumn{1}{l|}{6.29689453025e+30}  \\ \hline
\multicolumn{1}{|l|}{PB} & \multicolumn{1}{l|}{125.232526939} & \multicolumn{1}{l|}{6.39885542365e-28}  & \multicolumn{1}{l|}{1} & \multicolumn{1}{l|}{1.56277948757e+27}  \\ \hline
\multicolumn{1}{|l|}{PI} & \multicolumn{1}{l|}{105.735798465} & \multicolumn{1}{l|}{1.09588023355e-23}  & \multicolumn{1}{l|}{1} & \multicolumn{1}{l|}{9.12508474362e+22}  \\ \hline
\multicolumn{1}{|l|}{RN} & \multicolumn{1}{l|}{103.807827371} & \multicolumn{1}{l|}{2.87353636201e-23}  & \multicolumn{1}{l|}{1} & \multicolumn{1}{l|}{3.48003252446e+22}  \\ \hline
\multicolumn{1}{|l|}{RO} & \multicolumn{1}{l|}{92.3868106145} & \multicolumn{1}{l|}{8.6787858999e-21}   & \multicolumn{1}{l|}{1} & \multicolumn{1}{l|}{1.1522348996e+20}   \\ \hline
\multicolumn{1}{|l|}{RR} & \multicolumn{1}{l|}{83.7990976172} & \multicolumn{1}{l|}{6.35707240879e-19}  & \multicolumn{1}{l|}{1} & \multicolumn{1}{l|}{1.57305114005e+18}  \\ \hline
\multicolumn{1}{|l|}{SC} & \multicolumn{1}{l|}{130.020826581} & \multicolumn{1}{l|}{5.83896996879e-29}  & \multicolumn{1}{l|}{1} & \multicolumn{1}{l|}{1.71263083274e+28}  \\ \hline
\multicolumn{1}{|l|}{SE} & \multicolumn{1}{l|}{72.7348067195} & \multicolumn{1}{l|}{1.60633972519e-16}  & \multicolumn{1}{l|}{1} & \multicolumn{1}{l|}{6.22533318649e+15}  \\ \hline
\multicolumn{1}{|l|}{TO} & \multicolumn{1}{l|}{55.1303103352} & \multicolumn{1}{l|}{1.06808352921e-12}  & \multicolumn{1}{l|}{1} & \multicolumn{1}{l|}{936256362587}       \\ \hline
\multicolumn{1}{|l|}{BA} & \multicolumn{1}{l|}{342.184328822} & \multicolumn{1}{l|}{4.96154688314e-75}  & \multicolumn{1}{l|}{1} & \multicolumn{1}{l|}{2.01550045491e+74}  \\ \hline
\multicolumn{1}{|l|}{MG} & \multicolumn{1}{l|}{777.261043094} & \multicolumn{1}{l|}{1.65923917468e-169} & \multicolumn{1}{l|}{1} & \multicolumn{1}{l|}{6.02685866668e+168} \\ \hline
\multicolumn{1}{|l|}{SP} & \multicolumn{1}{l|}{749.737824971} & \multicolumn{1}{l|}{1.57217132373e-163} & \multicolumn{1}{l|}{1} & \multicolumn{1}{l|}{6.36062994474e+162} \\ \hline
\multicolumn{1}{|l|}{RJ} & \multicolumn{1}{l|}{901.70236979}  & \multicolumn{1}{l|}{1.57695115134e-196} & \multicolumn{1}{l|}{1} & \multicolumn{1}{l|}{6.34135051776e+195} \\ \hline
\multicolumn{1}{|l|}{RS} & \multicolumn{1}{l|}{451.109474546} & \multicolumn{1}{l|}{1.10362679252e-98}  & \multicolumn{1}{l|}{1} & \multicolumn{1}{l|}{9.06103409936e+97}  \\ \hline
\multicolumn{1}{|l|}{PE} & \multicolumn{1}{l|}{142.356671451} & \multicolumn{1}{l|}{1.22360590247e-31}  & \multicolumn{1}{l|}{1} & \multicolumn{1}{l|}{8.1725660033e+30}   \\ \hline
\multicolumn{1}{|l|}{PR} & \multicolumn{1}{l|}{334.950626527} & \multicolumn{1}{l|}{1.84669679188e-73}  & \multicolumn{1}{l|}{1} & \multicolumn{1}{l|}{5.41507411718e+72}  \\ \hline
\end{tabular}
\centering
\caption{Statistical comparison between the models. We used the Akaike's information criterion (AIC) to compare the two models: WOC (without competition) and WCB (with competition). The AIC lets us determine which model is more likely to  describe correctly the data and quantify by calculating the probabilities and an evidence radio. The probability column shows the likelihood of each model to be the most correctly. The evidence radio is the fraction of Probability WC by Probability WOC, which means how many times model WC is likely to be correct than model WOC.}
\label{AIC_Fed_Dep}
\end{table*}

\begin{table*}[]
\begin{tabular}{lllll}
\cline{2-5}
                        & \multicolumn{4}{c}{States deputies}                                                                                                                     \\ \hline
\multicolumn{1}{|l|}{State} & \multicolumn{1}{|l|}{$\Delta$ AIC} & \multicolumn{1}{|l|}{Probability A}              & \multicolumn{1}{|l|}{Probability B}            & \multicolumn{1}{|l|}{Evidence radio}                         \\ \hline
\multicolumn{1}{|l|}{AC} & \multicolumn{1}{l|}{576.061906458} & \multicolumn{1}{l|}{8.12356005108e-126} & \multicolumn{1}{l|}{1} & \multicolumn{1}{l|}{1.23098739187e+125} \\ \hline
\multicolumn{1}{|l|}{AL} & \multicolumn{1}{l|}{238.628928458} & \multicolumn{1}{l|}{1.52190160204e-52}  & \multicolumn{1}{l|}{1} & \multicolumn{1}{l|}{6.57072703427e+51}  \\ \hline
\multicolumn{1}{|l|}{AM} & \multicolumn{1}{l|}{682.650418552} & \multicolumn{1}{l|}{5.81226058412e-149} & \multicolumn{1}{l|}{1} & \multicolumn{1}{l|}{1.72050097467e+148} \\ \hline
\multicolumn{1}{|l|}{AP} & \multicolumn{1}{l|}{358.738756255} & \multicolumn{1}{l|}{1.26144655989e-78}  & \multicolumn{1}{l|}{1} & \multicolumn{1}{l|}{7.92740677087e+77}  \\ \hline
\multicolumn{1}{|l|}{CE} & \multicolumn{1}{l|}{420.054263752} & \multicolumn{1}{l|}{6.11470593755e-92}  & \multicolumn{1}{l|}{1} & \multicolumn{1}{l|}{1.63540162064e+91}  \\ \hline
\multicolumn{1}{|l|}{ES} & \multicolumn{1}{l|}{480.640448515} & \multicolumn{1}{l|}{4.26827816914e-105} & \multicolumn{1}{l|}{1} & \multicolumn{1}{l|}{2.34286510947e+104} \\ \hline
\multicolumn{1}{|l|}{GO} & \multicolumn{1}{l|}{989.587809594} & \multicolumn{1}{l|}{1.29938385781e-215} & \multicolumn{1}{l|}{1} & \multicolumn{1}{l|}{7.69595523285e+214} \\ \hline
\multicolumn{1}{|l|}{MA} & \multicolumn{1}{l|}{519.730902297} & \multicolumn{1}{l|}{1.38633608904e-113} & \multicolumn{1}{l|}{1} & \multicolumn{1}{l|}{7.21325808299e+112} \\ \hline
\multicolumn{1}{|l|}{MS} & \multicolumn{1}{l|}{439.886900022} & \multicolumn{1}{l|}{3.01837593293e-96}  & \multicolumn{1}{l|}{1} & \multicolumn{1}{l|}{3.31303993347e+95}  \\ \hline
\multicolumn{1}{|l|}{MT} & \multicolumn{1}{l|}{310.209118004} & \multicolumn{1}{l|}{4.35457632874e-68}  & \multicolumn{1}{l|}{1} & \multicolumn{1}{l|}{2.29643465749e+67}  \\ \hline
\multicolumn{1}{|l|}{PA} & \multicolumn{1}{l|}{685.433108646} & \multicolumn{1}{l|}{1.44574467234e-149} & \multicolumn{1}{l|}{1} & \multicolumn{1}{l|}{6.9168506662e+148}  \\ \hline
\multicolumn{1}{|l|}{PB} & \multicolumn{1}{l|}{400.461973128} & \multicolumn{1}{l|}{1.09846804884e-87}  & \multicolumn{1}{l|}{1} & \multicolumn{1}{l|}{9.10358750133e+86}  \\ \hline
\multicolumn{1}{|l|}{PI} & \multicolumn{1}{l|}{189.012621263} & \multicolumn{1}{l|}{9.0454627114e-42}   & \multicolumn{1}{l|}{1} & \multicolumn{1}{l|}{1.10552664016e+41}  \\ \hline
\multicolumn{1}{|l|}{RN} & \multicolumn{1}{l|}{249.978331952} & \multicolumn{1}{l|}{5.2226980642e-55}   & \multicolumn{1}{l|}{1} & \multicolumn{1}{l|}{1.91471915035e+54}  \\ \hline
\multicolumn{1}{|l|}{RO} & \multicolumn{1}{l|}{482.146906385} & \multicolumn{1}{l|}{2.00969219136e-105} & \multicolumn{1}{l|}{1} & \multicolumn{1}{l|}{4.97588637852e+104} \\ \hline
\multicolumn{1}{|l|}{RR} & \multicolumn{1}{l|}{370.038828903} & \multicolumn{1}{l|}{4.4369982636e-81}   & \multicolumn{1}{l|}{1} & \multicolumn{1}{l|}{2.25377595525e+80}  \\ \hline
\multicolumn{1}{|l|}{SC} & \multicolumn{1}{l|}{462.924309949} & \multicolumn{1}{l|}{3.00098154818e-101} & \multicolumn{1}{l|}{1} & \multicolumn{1}{l|}{3.33224308096e+100} \\ \hline
\multicolumn{1}{|l|}{SE} & \multicolumn{1}{l|}{139.093088068} & \multicolumn{1}{l|}{6.2563306112e-31}   & \multicolumn{1}{l|}{1} & \multicolumn{1}{l|}{1.59838100341e+30}  \\ \hline
\multicolumn{1}{|l|}{TO} & \multicolumn{1}{l|}{282.919902956} & \multicolumn{1}{l|}{3.67048676832e-62}  & \multicolumn{1}{l|}{1} & \multicolumn{1}{l|}{2.72443428656e+61}  \\ \hline
\multicolumn{1}{|l|}{BA} & \multicolumn{1}{l|}{642.920743716} & \multicolumn{1}{l|}{2.46339667887e-140} & \multicolumn{1}{l|}{1} & \multicolumn{1}{l|}{4.0594355289e+139}  \\ \hline
\multicolumn{1}{|l|}{MG} & \multicolumn{1}{l|}{1450.44716375} & \multicolumn{1}{l|}{1.09496501832e-315} & \multicolumn{1}{l|}{1} & \multicolumn{1}{l|}{inf}                \\ \hline
\multicolumn{1}{|l|}{SP} & \multicolumn{1}{l|}{2129.42600533} & \multicolumn{1}{l|}{0}                  & \multicolumn{1}{l|}{1} & \multicolumn{1}{l|}{inf}                \\ \hline
\multicolumn{1}{|l|}{RJ} & \multicolumn{1}{l|}{1694.58149782} & \multicolumn{1}{l|}{0}                  & \multicolumn{1}{l|}{1} & \multicolumn{1}{l|}{inf}                \\ \hline
\multicolumn{1}{|l|}{RS} & \multicolumn{1}{l|}{618.918508849} & \multicolumn{1}{l|}{4.01377875167e-135} & \multicolumn{1}{l|}{1} & \multicolumn{1}{l|}{2.49141784306e+134} \\ \hline
\multicolumn{1}{|l|}{PE} & \multicolumn{1}{l|}{369.31393498}  & \multicolumn{1}{l|}{6.37526107172e-81}  & \multicolumn{1}{l|}{1} & \multicolumn{1}{l|}{1.56856321451e+80}  \\ \hline
\multicolumn{1}{|l|}{PR} & \multicolumn{1}{l|}{784.782590509} & \multicolumn{1}{l|}{3.8603414867e-171}  & \multicolumn{1}{l|}{1} & \multicolumn{1}{l|}{2.59044440354e+170} \\ \hline
\end{tabular}
\centering
\caption{Statistical comparison between the models. We used the Akaike's information criterion (AIC) to compare the two models: WC (without competition) and WOC (with competition). The AIC lets us determine which model is more likely to  describe correctly the data and quantify by calculating the probabilities and an evidence radio. The probability column shows the likelihood of each model to be the most correctly. The evidence radio is the fraction of Probability WC by Probability WOC, which means how many times model WC is likely to be correct than model WOC.  }
\label{AIC_Est_Dep}
\end{table*}
\newpage

\begin{thebibliography}{11}

\bibitem{NYtimes}
\url{http://elections.nytimes.com/2012/campaign-finance} (Accessed:
  2017-10-11).

\bibitem{Stratmann05}
Stratmann T (2005) Some talk: Money in politics. a (partial) review of the
  literature.
\newblock {\em Policy Challenges and Political Responses} 124(1-2):135--156.

\bibitem{Holbrook96}
Holbrook T (1996) {\em Do campaigns matter?}
\newblock (Sage, London).

\bibitem{Johnston06}
Johnston RG, Brady HE (2009) {\em Capturing campaign effects}.
\newblock (University of Michigan Press, Ann Arbor).

\bibitem{Jacobson1978}
Jacobson GC (1978) The effects of campaign spending in congressional elections.
\newblock {\em Am Polit Sci Rev} 72(2):469--491.

\bibitem{GERBER04}
Gerber AS (2004) Does campaign spending work? field experiments provide
  evidence and suggest new theory.
\newblock {\em Am Behav Sci} 47(5):541--574.

\bibitem{Johnston08}
Johnston R, Pattie C (2008) How much does a vote cost? incumbency and the
  impact of campaign spending at english general elections.
\newblock {\em J Elect Public Opin Parties} 18(2):129--152.

\bibitem{Erikson2000}
Erikson RS, Palfrey TR (2000) Equilibria in campaign spending games: Theory and
  data.
\newblock {\em Am Polit Sci Rev} 94(3):595--609.

\bibitem{Hillygus03}
Hillygus DS, Jackman S (2003) Voter decision making in election 2000: Campaign
  effects, partisan activation, and the clinton legacy.
\newblock {\em Am J Pol Sci} 47(4):583--596.

\bibitem{Lazarsfeld1944}
Lazarsfeld PF, Berelson B, Gaudet H (1948) {\em The peoples choice: how the
  voter makes up his mind in a presidential campaign.}
\newblock (Columbia University Press, New York).

\bibitem{Finkel1993}
Finkel SE (1993) Reexamining the" minimal effects" model in recent presidential
  campaigns.
\newblock {\em J Polit} 55(1):1--21.

\bibitem{Krasno1988}
Krasno JS, Green DP (1988) Preempting quality challengers in house elections.
\newblock {\em J Polit} 50(4):920--936.

\bibitem{West2013}
West DM (2013) {\em Air wars: Television advertising and social media in
  election campaigns, 1952-2012}.
\newblock (Sage, London).

\bibitem{Esser2004}
Esser F, Pfetsch B (2004) {\em Comparing political communication: Theories,
  cases, and challenges}.
\newblock (Cambridge University Press, New York).

\bibitem{McAfee1995}
McAfee RP, McMillan J (1995) Organizational diseconomies of scale.
\newblock {\em J Econ Manag Strategy} 4(3):399--426.

\bibitem{Ringelmann1913}
Ringlemann M (1913) Recherches sur les moteurs anim{\'e}s: Travail de l’homme
  in {\em Annales de l’Institut National Agronomique}.
\newblock Vol.{}~12, pp. 1--40.

\bibitem{Kitsak2010}
Kitsak M, et~al. (2010) Identification of influential spreaders in complex
  networks.
\newblock {\em Nat Phys} 6:888--893.

\bibitem{Morone2015}
Morone F, Makse HA (2015) Influence maximization in complex networks through
  optimal percolation.
\newblock {\em Nature} 524:65--68.

\bibitem{TSE}
\url{http://www.tse.gov.br/} (Accessed: 2017-10-11).

\bibitem{Milgram67}
Milgram S (1967) The small world problem.
\newblock {\em Psychol Today} 1:61--67.

\bibitem{Barabasi2016}
Barab{\'a}si AL (2016) {\em Network science}.
\newblock (Cambridge university press, New York).

\bibitem{Costa99}
Costa~Filho R, Almeida M, Andrade J, Moreira J, , et~al. (1999) Scaling
  behavior in a proportional voting process.
\newblock {\em Phys Rev E} 60(1):1067.

\bibitem{Costa2003}
Costa~Filho R, Almeida M, Moreira J, Andrade J (2003) Brazilian elections:
  voting for a scaling democracy.
\newblock {\em Physica A} 322:698--700.

\bibitem{Moreira06}
Moreira AA, Paula DR, Costa~Filho RN, Andrade~Jr JS (2006) Competitive cluster
  growth in complex networks.
\newblock {\em Phys Rev E} 73(6):065101.

\bibitem{Loreto09}
Castellano C, Fortunato S, Loreto V (2009) Statistical physics of social
  dynamics.
\newblock {\em Rev Mod Phys} 81(2):591.

\bibitem{Calvao2015}
Calv{\~a}o AM, Crokidakis N, Anteneodo C (2015) Stylized facts in brazilian
  vote distributions.
\newblock {\em PloS one} 10(9):e0137732.

\bibitem{Fortunato2007}
Fortunato S, Castellano C (2007) Scaling and universality in proportional
  elections.
\newblock {\em Phys Rev Lett} 99(13):138701.

\bibitem{Jaynes1957}
Jaynes ET (1957) Information theory and statistical mechanics.
\newblock {\em Phys Rev} 106(4):620.

\bibitem{Klimek2012}
Klimek P, Yegorov Y, Hanel R, Thurner S (2012) Statistical detection of
  systematic election irregularities.
\newblock {\em Proc Natl Acad Sci} 109(41):16469--16473.

\bibitem{Borghesi2010}
Borghesi C, Bouchaud JP (2010) Spatial correlations in vote statistics: a
  diffusive field model for decision-making.
\newblock {\em Eur Phys J B} 75(3):395--404.

\bibitem{Nuno2010}
Ara{\'u}jo NA, Andrade~Jr JS, Herrmann HJ (2010) Tactical voting in plurality
  elections.
\newblock {\em PLoS One} 5(9):e12446.

\bibitem{Borghesi2012}
Borghesi C, Raynal JC, Bouchaud JP (2012) Election turnout statistics in many
  countries: similarities, differences, and a diffusive field model for
  decision-making.
\newblock {\em PloS one} 7(5):e36289.

\bibitem{Andresen2008}
Andresen CA, Hansen HF, Hansen A, Vasconcelos GL, Andrade~Jr JS (2008)
  Correlations between political party size and voter memory: A statistical
  analysis of opinion polls.
\newblock {\em Int J Mod Phys C} 19(11):1647--1657.

\bibitem{Mantovani2011}
Mantovani M, Ribeiro H, Moro M, Picoli~Jr S, Mendes R (2011) Scaling laws and
  universality in the choice of election candidates.
\newblock {\em Europhys Lett} 96(4):48001.

\bibitem{Fortunato2013}
Chatterjee A, Mitrovi{\'c} M, Fortunato S (2013) Universality in voting
  behavior: an empirical analysis.
\newblock {\em Sci Rep} 3:1049.

\end{thebibliography}

\begin{thebibliography}{11}
\bibitem{S_TSE} http://www.tse.gov.br/
\bibitem{S_Motulsky2004} Motulskuy H, Christopoulos A (2004)
Fitting models to biological data using linear and nonlinear 
regression: a practical guide to curve fitting (Oxford University Press)
\bibitem{S_Jaynes1957} Jaynes ET (1957) Information theory and 
statistical mechanics. \emph{Phys. Rev.} 106(4):620.
\end{thebibliography}
\end{document}